\documentclass[fleqn,usenatbib]{mnras}

\usepackage{newtxtext,newtxmath}

\usepackage[T1]{fontenc}

\DeclareRobustCommand{\VAN}[3]{#2}
\let\VANthebibliography\thebibliography
\def\thebibliography{\DeclareRobustCommand{\VAN}[3]{##3}\VANthebibliography}

\usepackage{graphicx}	
\usepackage{amsmath}	
\usepackage{multirow}
\usepackage{textcomp}
\usepackage{placeins}
\usepackage{float}
\newcommand{\lya}{Ly$\alpha$}
\newcommand{\lyb}{Ly$\beta$}

\newcommand{\hi}{\ion{H}{i}}

\newcommand{\taueff}{$\tau_{\rm eff}$}

\newcommand{\mfp}{$\lambda_{\rm mfp}$}

\newcommand{\Req}{$R_{\rm eq}$}

\title[Higher-order Lyman-series absorption at $z=5$--6]{Constraining the end of reionization with 
high-order Lyman-series absorption }

\author[ Hashemi et al.]{
Seyedazim Hashemi,$^{1}$\thanks{E-mail: seyedazim.hashemi@email.ucr.edu}
George D. Becker,$^{1}$
Yongda Zhu,$^{2}$ 
James S. Bolton,$^{3}$
Hui Hong$^{1}$
\\
$^{1}$Department of Physics \& Astronomy, University of California, Riverside, CA, 92521, USA\\
$^{2}$Steward Observatory, University of Arizona, 933 N Cherry Ave, Tucson, AZ 85721, USA\\
$^{3}$School of Physics and Astronomy, University of Nottingham, University Park, Nottingham, NG7 2RD, UK\\
}

\date{Accepted XXX. Received YYY; in original form ZZZ}

\pubyear{\the\year{}}

\begin{document}
\label{firstpage}
\pagerange{\pageref{firstpage}--\pageref{lastpage}}
\maketitle

\begin{abstract}
Both \lya\ forest and Lyman-continuum (LyC) observations favor a relatively late end to reionization, while leaving open the question of whether current models accurately capture the full neutral hydrogen column-density distribution of the intergalactic medium (IGM). Higher-order Lyman-series (LyS) absorption provides a complementary probe of this distribution, remaining sensitive to progressively higher \hi\ column densities. We use four quasar composite spectra at $z = 5.08$, 5.32, 5.64, and 5.93, constructed from 100 quasars observed with Keck/LRIS, Keck/ESI, Gemini/GMOS, and VLT/X-Shooter. We then generate forward-modeled composite spectra from the Sherwood simulation, which assumes an early and spatially homogeneous reionization, and from four Sherwood--Relics radiative-transfer models that incorporate patchy reionization with different hydrogen reionization completion redshifts. The cumulative higher-order LyS transmission in the observed composites is then compared to that predicted by each model. We find that the observed transmission is best reproduced by the Sherwood--Relics model in which reionization completes at $z\sim5.3$. The same model also reproduces recent  \mfp\ measurements based on LyC transmission alone, further supporting this late-reionization scenario. Compared with models that reionize substantially earlier, the enhanced higher-order LyS absorption in the preferred model arises from differences in the gas densities, gas temperatures, and residual neutral regions, with the thermal state dominating at lower redshifts and residual neutral regions at higher redshifts. Our results establish higher-order hydrogen LyS absorption as an effective probe of the \hi\ absorber population and late stages of cosmic reionization.
\end{abstract}
\begin{keywords}
intergalactic medium - quasars: absorption lines - cosmology: observations - dark ages, reionization, first stars - large-scale structure of the Universe
\end{keywords}



\section{Introduction}
The Epoch of Reionization represents the last major phase transition of baryonic matter in the Universe. During this period, ionizing radiation produced by the first generations of stars, galaxies, and accreting black holes transformed the intergalactic medium (IGM) from almost fully neutral to highly ionized. Despite substantial observational and theoretical progress, important uncertainties remain in the detailed timing and topology of the final stages of reionization (see e.g.\ \citealt{Becker2015, Wise2019, Robertson2022} for reviews).

A diverse range of observations, including cosmic microwave background (CMB) polarization, redshifted 21 cm measurements, quasar and galaxy \lya\ damping-wing analyses, and \lya\ emitter luminosity functions, have provided constraints on the timing and evolution of cosmic reionization, with current measurements generally favoring a midpoint near $z\sim7$--8 and an increasingly neutral IGM toward higher redshift (e.g.\ \citealt{ Banados2018,Davies2018b, Greig2019,planck2020,HERA2023,Umeda2024,Kageura2025,Mertens2025, Munshi2025, Ceccotti2025,Nunhokee2025,Chen2026,Umeda2026}). At lower redshifts, observations of the \lya\ forest in quasars at $z\gtrsim5$--6 indicate that the ionization state of the IGM near the end of reionization is highly inhomogeneous (e.g.\ \citealt{Becker2015b, Bosman2018, Eilers2018, Yang2020, Zhu2021, Bosman2022}). While some models explain the large \lya\ opacity fluctuations primarily through spatial variations in the ionizing background and the mean free path of ionizing photons (\mfp) within an already highly ionized IGM (e.g.\ \citealt{Davies2016, Maity2026}), models involving patchy and incomplete reionization often provide a natural explanation (e.g.\ \citealt{Becker2018, Kulkarni2019, Keating2020a, Nasir2020, Choudhury2021, Qin2021, Maksora2026, Conaboy2026}). This picture is further supported by dark gaps in the \lyb\ forest (\citealt{Zhu2022}), dark pixels in the \lya, \lyb, and Ly$\gamma$ forests (e.g.,\ \citealt{Maity2026, Davies2026}), the rapid evolution of \mfp\ over $5 \lesssim z \lesssim 6$ \citep{Becker2021, Zhu2023, Gaikwad2023, Davies2024, Satyavolu2024}, the evolution of  metal absorbers near $z\sim6$ (e.g.\ \citealt{Becker2019,Cooper2019, DOdorico2022}),  as well as the association of large-scale galaxy underdensities with extended dark \lya\ troughs (\citealt{Becker2018, Kashino2020, Christenson2023, Conaboy2026, Zhu2026}).

At redshifts approaching $z\sim6$, \lya\ absorption becomes increasingly saturated as the neutral hydrogen fraction of the IGM increases. Owing to the large oscillator strength of the \lya\ transition, even a small neutral fraction ($x_{\mathrm{HI}}\gtrsim10^{-4}$) can produce nearly complete absorption \citep{Gunn1965}, limiting the sensitivity of the transmitted \lya\ flux to further increases in neutral hydrogen. Previous studies have therefore extended the analysis to \lyb\ and Ly$\gamma$ absorption (e.g.\ \citealt{Davies2018a,Eilers2019,Keating2020b,Zhu2022,Davies2026}), whose smaller oscillator strengths partially alleviate this saturation. However, their optical depths are smaller than that of \lya\ by factors of approximately $6$ and $18$, respectively, and therefore provide only a limited extension to the range of \hi\ column densities that can be probed before saturation. At the same time, measurements of the ionizing photon mean free path are primarily sensitive to absorbers with column densities roughly three to four orders of magnitude higher than those at which \lya\ absorption becomes saturated. Late-ending reionization models are supported by both the \lya\ forest, which primarily probes gas near or below the cosmic mean density and is sensitive to the overall ionization, and Lyman-continuum (LyC) measurements~\citep{Becker2021, Zhu2023}, which are sensitive to the abundance of much higher-column-density absorbers. These observables therefore provide complementary constraints on two critical aspects of reionization. However, it remains unclear whether current models self-consistently reproduce the absorber population responsible for the observed \mfp\ and its rapid evolution.

Higher-order Lyman-series (LyS) absorption provides a natural way to connect these regimes. The optical depths of LyS transitions scale as $\tau_n\propto N_{\rm HI}f_n\lambda_n$. Because the oscillator strengths of successive transitions decrease rapidly with increasing principal quantum number, progressively higher-order lines remain sensitive to increasingly large \hi\ column densities after lower-order transitions have saturated.
 For example, while \lya\ ($n=2$) has $\lambda\simeq1216$~\AA\ and $f\simeq0.42$, a transition with $n\sim30$ lies near $\lambda\simeq913$~\AA\ and has $f\sim6\times10^{-5}$, reducing its optical depth by nearly four orders of magnitude at fixed $N_{\rm HI}$. In the spectra of quasars at $z\sim5$--6, the region between \lya\ and the Lyman limit contains the cumulative absorption from multiple LyS transitions arising at different foreground redshifts, with progressively more transitions contributing toward the Lyman limit. This cumulative LyS absorption therefore carries sensitivity to a broad range of \hi\ column densities, bridging the diffuse IGM traced by the lower-order Lyman forest and the higher-column-density absorber population that strongly influences the \mfp.

In this work, we model the cumulative LyS transmission including transitions up to \(n=1000\). Using  Sherwood \citep{Bolton2017} and Sherwood-Relics \citep{Puchwein2023} simulation suites, which span a wide range of reionization histories, we generate synthetic spectra spanning the full rest-frame wavelength range blueward of \lya\ emission, approximately 820--1215~\AA, and compare their cumulative LyS transmission with that measured from observed quasar composite spectra. In particular, we examine scenarios that differ in the timing of reionization completion, the photoionization rate ($\Gamma_{\rm HI}$), and the thermal state of the IGM. This approach enables us to investigate how different reionization histories affect the combined LyS absorption, providing a new observational avenue for constraining the final stages of reionization.

The rest of the paper is organized as follows. In Section~\ref{sec:data}, we describe the observational data and simulations used in this work. Section~\ref{sec:transmission} presents our methodology and the resulting LyS transmission measurements for the different simulation models. In Section~\ref{sec:discussion}, we discuss the physical interpretation of these measurements and their implications for the late stages of  reionization. Finally, we summarize our main conclusions in Section~\ref{sec:summary}. Throughout this paper, we adopt $\Omega_{\rm m}=0.308$, $\Omega_{\Lambda}=0.692$, and $h=0.678$~\citep{Planck2014}.

\section{Quasar Composites }
\label{sec:data}

\begin{figure*}
    \centering
    \includegraphics[width=\linewidth]{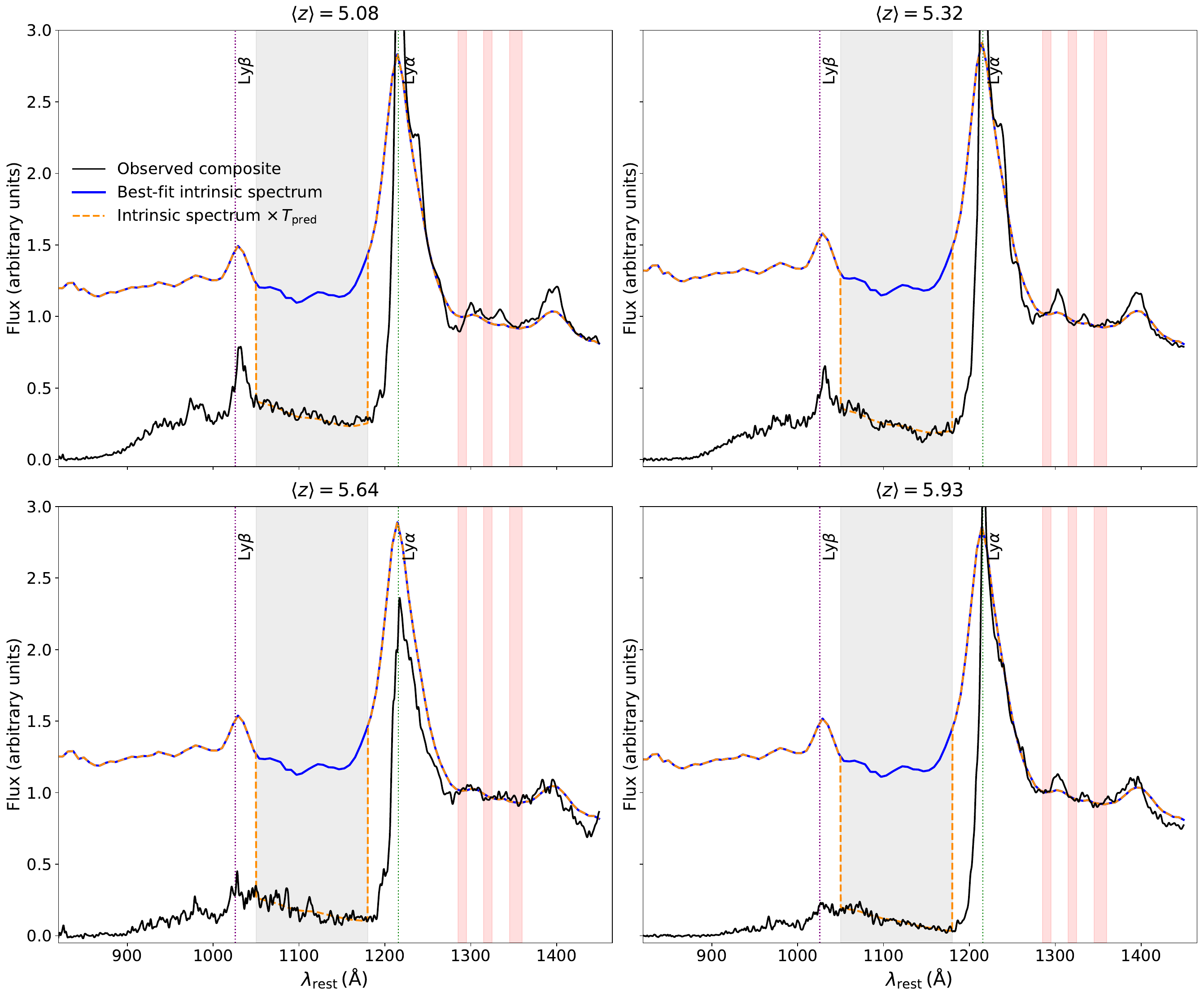}
    \caption{Quasar composite spectra used in this work. The  observed composite spectra are shown in black, while the best-fitting intrinsic spectra, obtained by rescaling and applying a power-law tilt to the transmission-corrected low-redshift quasar composite of \citet{Lusso2015}, are shown in blue. The orange dashed curves show the best-fitting intrinsic spectra multiplied by the expected mean IGM transmission within the \lya\ forest fitting interval. See Section~\ref{sec:data} for details of the continuum fitting. The vertical dotted lines mark the rest-frame wavelengths of \lyb\ and \lya. The light-red shaded regions and the gray-shaded \lya\ forest interval indicate the wavelength ranges used for continuum fitting.}
    \label{fig:composites}
\end{figure*}

\begin{figure*}
    \centering
    \includegraphics[width=\linewidth]{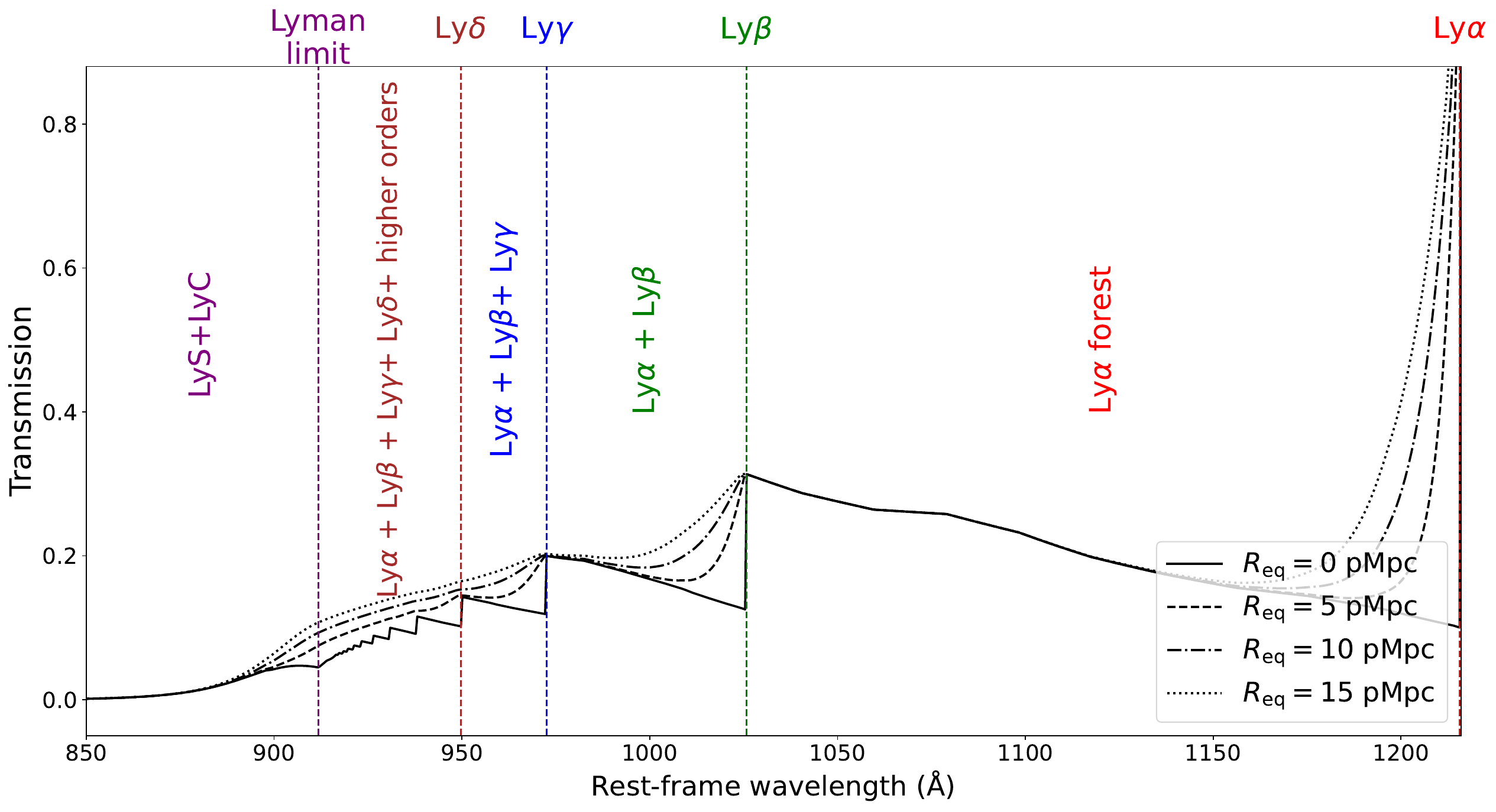}
    \caption{
Illustration of the transmission predicted by  Sherwood--Relics simulation for a quasar at $z=5.32$. The vertical dashed lines mark the wavelengths of the \lya, \lyb, Ly$\gamma$, Ly$\delta$, and Lyman-limit transitions. The labeled wavelength intervals indicate the dominant absorption regimes blueward of the \lya\ emission line, progressing from absorption by the \lya\ forest alone to the cumulative contribution of higher-order LyS transitions and, below the Lyman limit, LyC absorption. The different line styles illustrate the effect of varying the  quasar proximity-zone size, $R_{\rm eq}$, on the transmitted flux.}
    \label{fig:illustration}
\end{figure*}

Our analysis is based on modified versions of the quasar composite spectra presented by \citet{Zhu2023}. The original composites were constructed from 97 high-quality quasar spectra spanning the redshift range $4.9<z<6.1$, obtained with multiple spectrographs, including Keck/LRIS, Gemini/GMOS, Keck/ESI, and VLT/X-Shooter. Individual quasar spectra were first shifted to the rest frame and then normalized by the continuum flux measured over the rest-frame wavelength range $1270<\lambda<1380$~\AA, which is relatively free of emission lines. The normalized spectra were subsequently combined to produce composite spectra in redshift bins of width $\Delta z=0.3$. For the present work, we modify two of these composites, as described below.

The original $\langle z \rangle = 5.31$ composite was modified by including an ESI spectrum of quasar J0120+2147, which was not part of the sample used by \citet{Zhu2023} but has since been presented by \citet{Zhu2026}. J0120+2147 has a $ [\mathrm{C\,II}]\text{-based}$ redshift of $z_{\rm [CII]} = 5.4305$ and an absolute magnitude of $M_{1450}=-26.5$. The addition of J0120+2147 increases the number of quasars contributing to the composite from 26 to 27 and shifts the mean redshift from $\langle z \rangle = 5.31$ to $\langle z \rangle = 5.32$.

The original $\langle z \rangle = 5.65$ composite was modified by including ESI spectra of two additional quasars, J0056+2241 and J1016+2541, which were not part of the sample used by \citet{Zhu2023} but have also since been presented by \citet{Zhu2026}. J0056+2241 has a systemic redshift of $z_{\rm [CII]} = 5.5218$ and an absolute magnitude of $M_{1450}=-26.8$, while J1016+2541 has $z_{\rm [CII]} = 5.6797$ and $M_{1450}=-27.7$. The addition of these two quasars increases the number of spectra contributing to the composite from 9 to 11 and shifts the mean redshift from $\langle z \rangle = 5.65$ to $\langle z \rangle=5.64$.

The $\langle z \rangle = 5.08$ and $\langle z \rangle = 5.93$ composites were adopted directly from \citet{Zhu2023}. The $\langle z \rangle = 5.08$ composite therefore consists of the original 44 quasars, while the $\langle z \rangle = 5.93$ composite consists of the original 18 quasars.

In this work we analyze the continuum-normalized LyS transmission.  To estimate the mean intrinsic continua of the composites, we use the transmission-corrected $z \simeq 2.4$ quasar composite spectrum presented by \cite{Lusso2015}. As illustrated in Fig.~\ref{fig:composites}, the mean continuum for each composite is modeled as a rescaled and slightly tilted version of the Lusso et al. template,
\begin{equation}
F_{\rm int}(\lambda)=
A\,F_{\rm Lusso}(\lambda)
\left(\frac{\lambda}{1285\,\mbox{\AA}}\right)^{\beta},
\end{equation}
where $A$ is a multiplicative normalization and $\beta$ accounts for small differences in continuum slope between the observed composite and the intrinsic template. These parameters are determined by simultaneously fitting the observed composite in continuum windows largely free of strong emission features ($1285$--$1295$~\AA, $1315$--$1325$~\AA, and $1345$--$1360$~\AA). We also include the \lya\ forest from $1050$--$1180$~\AA\ in our fit after dividing the composite values by the expected global  mean IGM transmission,
$$
T_{\rm pred}(\lambda)
=
\exp\!\left[-\tau_{\rm eff}^{\rm Ly\alpha}(z_{\rm abs})\right],
$$
where the absorber redshift is given by
$$
z_{\rm abs}
=
\frac{\lambda(1+z_{\rm QSO})}
{\lambda_{\rm Ly\alpha}}
-1.
$$
Including the \lya\ forest implicitly assumes that the mean \lya\ transmission in our composites is close to the global mean. This is justified by the typically large number of individual spectra contributing to each composite. It also assumes that variations in the mean \lya\ flux of our composites are at least equally likely to arise from variations in the underlying continua of the objects as from variations in the average IGM \lya\ transmission, which is reasonable given the object-to-object variation in red-side continua. We have tested omitting the \lya\ forest from our continuum fits and find that it does not significantly alter our conclusions. Nevertheless, we emphasize that our analysis is best interpreted as a relative comparison between the \lya\ and LyS transmission in each composite.

The resulting absorbed model,
\begin{equation}
F_{\rm model}(\lambda)
=
F_{\rm int}(\lambda)\,
T_{\rm pred}(\lambda),
\end{equation}
is fitted directly to each composite, allowing the red-side continuum and the \lya\ forest to jointly constrain the intrinsic spectrum. The resulting best-fitting intrinsic spectrum is then used to estimate the average IGM transmission blueward of \lya\ as
$$
T_{\rm IGM}(\lambda)
=
\frac{F_{\rm obs}(\lambda)}
{F_{\rm int}(\lambda)}.
$$
Fig.~\ref{fig:composites} shows the resulting best-fitting intrinsic continua and the corresponding absorbed models.
For clarity, the composite spectra have been lightly smoothed with a Gaussian kernel of width $\sigma=2$ pixels (equivalent to $240~{\rm km~s^{-1}}$). We have also verified that fitting the intrinsic spectrum using only the \lya\ forest produces results consistent with our fiducial method. In addition, repeating the analysis using a simple power-law continuum instead of the Lusso et al. template does not alter our conclusions. Together, these tests suggest that our results are not highly sensitive to the specific continuum-fitting procedure adopted.

\section{Transmission Models}
\label{sec:transmission}

\subsection{Simulations}
\label{subsec:simulations}

 We  compare our empirically estimated transmission spectra blueward of \lya\ with the transmission predicted by the models.
To model the transmission blueward of \lya, we use two complementary simulation suites: Sherwood and Sherwood-Relics. 

The Sherwood simulations are cosmological hydrodynamical simulations performed using a modified version of the parallel Tree-PM smoothed particle hydrodynamics (SPH) code \textsc{P-Gadget-3}, an updated and extended version of the publicly available \textsc{Gadget-2} \citep{Springel2005}. Photoionization and photoheating of hydrogen and helium are computed using the spatially uniform ultraviolet background (UVB) model of \citet{Haardt2012}. The gas is assumed to be optically thin and in photoionization equilibrium, an approximation that is expected to be highly reliable for the post-reionization IGM, except in rare, highly overdense regions. In the \citet{Haardt2012} model, hydrogen reionization is completed rapidly at $z_{\rm r}=15$. In addition to having no neutral regions at $z<6$, this early reionization also produces more pressure smoothing and lower IGM temperatures at $z<6$ compared to models that reionized later. A comprehensive description of the Sherwood simulations is provided by \citet{Bolton2017}, to which we refer the interested reader for further details.

The Sherwood-Relics simulations extend the Sherwood simulation suite by incorporating a broader range of astrophysical models relevant to the high-redshift IGM (\citealt{Puchwein2023}). Specifically, they explore a wider range of reionization and thermal histories. As a result, these simulations provide a physically motivated framework for studying the impact of different reionization scenarios on the \lya\ forest and higher-order LyS absorption at $z>4$. To this end, they employ a new hybrid radiative transfer/cosmological hydrodynamical simulation technique that follows an inhomogeneous cosmic reionization process as well as the associated heating and pressure smoothing. The scheme uses radiation fields from post-processing radiative transfer simulations to photoionize and photoheat the IGM. These ionization and heating histories are then implemented in a subsequent cosmological hydrodynamical simulation that captures the hydrodynamic response to the heating. For a comprehensive description of the Sherwood-Relics simulations, we refer the reader to \citet{Puchwein2023}.

For both simulation suites, we use a $40\,h^{-1}\,\mathrm{cMpc}$ box with $2\times2048^3$ particles. For the Sherwood-Relics simulations,
we consider four reionization histories in which hydrogen reionization is completed at $z_{\rm r}=5.3$, 5.7, 6.0, and 6.6 (defined as the redshift at which the volume-averaged neutral hydrogen fraction falls below $10^{-3}$), corresponding to midpoint redshifts of $z_{\rm mid}=7.2$, 7.5, 7.3, and 8.0, respectively.  Throughout this work, we refer to these models as Relics53, Relics57, Relics60, and Relics66, respectively. A summary of the simulations used in this study is provided in Table~\ref{tab:simulations}.

Convergence of the model transmissions with box size and mass resolution was checked using available Sherwood runs, in which we varied the mass resolution and box size separately. The results indicate that we are well converged with mass resolution at our nominal value of $M_{\rm gas}=9.97\times10^{4}\,h^{-1}\,\mathrm{M}_{\odot}$, with only about a 4 per cent increase in transmission at 912~\AA\ when using a factor of 8 coarser mass resolution. We also find that the results obtained with our nominal box size of $40\,h^{-1}{\rm cMpc}$ underpredict the transmission by roughly 1 per cent compared to those from the $160\,h^{-1}{\rm cMpc}$ box at fixed resolution. This is presumably because the smaller box contains fewer rare, deep voids, which tend to increase the transmission. None of these small differences significantly changes the conclusions of this work.

The Sherwood and Sherwood–Relics simulations employ the quick-\lya\ approximation, in which gas particles with overdensity $\Delta>1000$ and temperature $T<10^5$ K are converted into collisionless star particles~\citep{Viel2004}. Consequently, these simulations may not accurately capture the contribution of high-density gas ($\Delta>1000$ and $T<10^5$ K) to the predicted LyS transmission. We test the sensitivity of our results to this approximation using two Sherwood simulations with different treatments of dense gas, as described in Appendix~\ref{sec:quick_lya_effects}.

\begin{table}
\centering
\caption{Summary of the simulations used in this work. Here, $z_{\rm r}$ denotes the redshift at which hydrogen reionization is completed.}
\label{tab:simulations}
\begin{tabular}{lccc}
\hline
Simulation & Box size & $z_{\rm r}$ & UVB model \\
           & [$h^{-1}\,\mathrm{cMpc}$] & & \\
\hline

Sherwood & 40 & 15  & Uniform~(\citealt{Haardt2012}) \\

Relics53 & 40 & 5.3 & Patchy~(\citealt{Puchwein2023}) \\

Relics57 & 40 & 5.7 & Patchy~(\citealt{Puchwein2023}) \\

Relics60 & 40 & 6.0 & Patchy~(\citealt{Puchwein2023}) \\

Relics66 & 40 & 6.6 & Patchy~(\citealt{Puchwein2023}) \\

\hline
\end{tabular}
\end{table}

\begin{figure*}
    \centering
    \includegraphics[width=\linewidth]{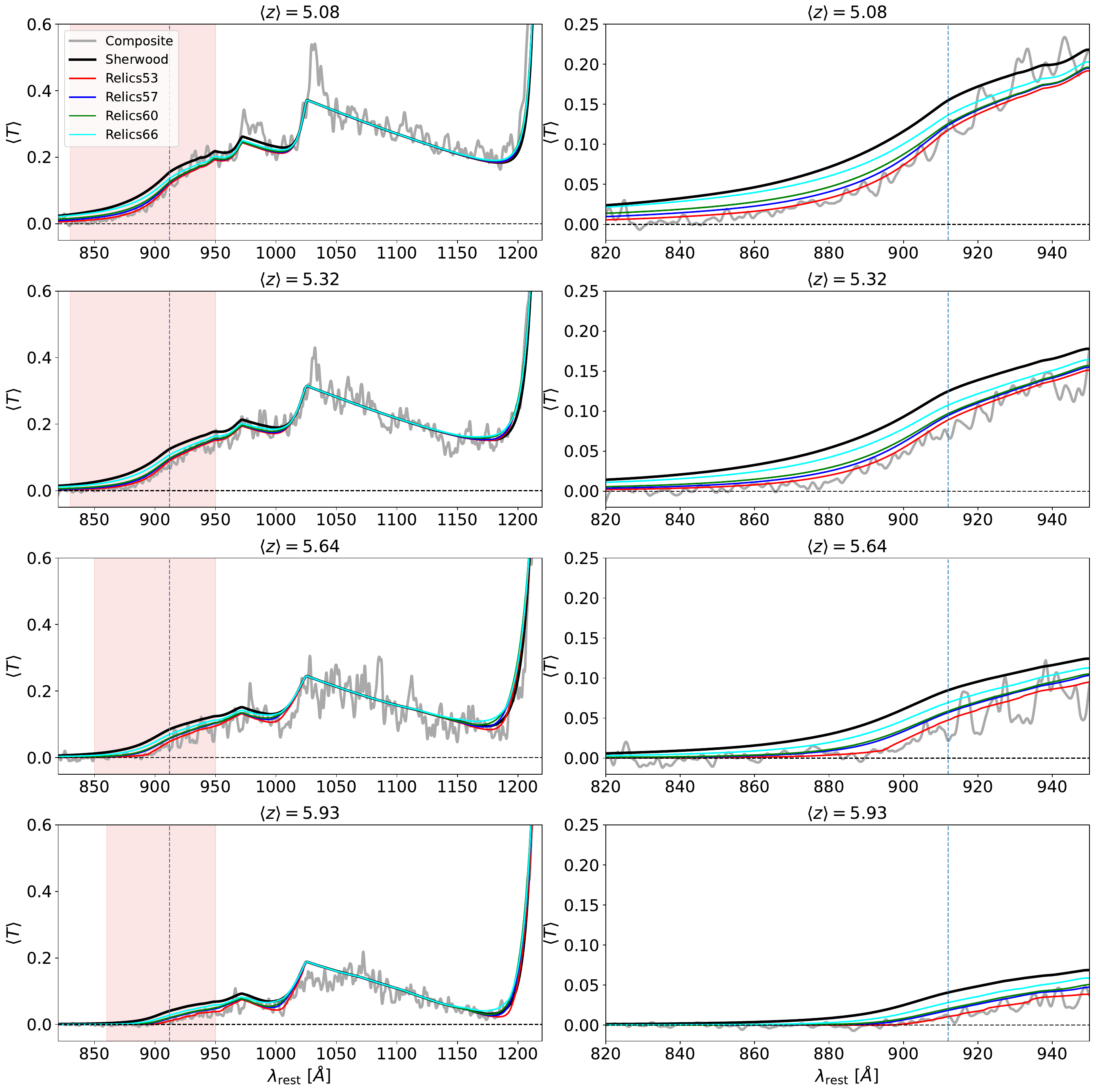}
    \caption{ Comparison between the observed quasar composite transmission spectra (gray) and the predictions of the Sherwood (black) and Sherwood-Relics models (red, blue, green, and cyan) at mean quasar redshifts $\langle z \rangle = 5.08$, 5.32, 5.64, and 5.93. The Sherwood-Relics models correspond to reionization histories ending at $z_{\rm r}=5.3$ (Relics53), 5.7 (Relics57), 6.0 (Relics60), and 6.6 (Relics66). The left column shows the predicted and observed transmission over the full rest-frame wavelength range considered in this work, while the right column highlights the region between 820 and 950~\AA, where the differences between the models are most pronounced. The vertical dashed line marks the Lyman limit at $\sim912$~\AA. The light-red shaded regions in the left panels indicate the wavelength ranges used for the $\chi^2$ analysis.}
    \label{fig:transmissions_zeta67}
\end{figure*}

\subsection{Formalism}
\label{subsec:methods}

To calculate the transmission at each observed wavelength in the composite spectra, it is necessary to account for absorption from all foreground LyS transitions and the Lyman continuum. Following \citet{Becker2021}, the cumulative effective optical depth from foreground LyS absorption is obtained by summing the contributions from all relevant transitions along the line of sight,

\begin{equation}
\tau_{\rm eff}^{\rm Lyman}(\lambda_{\rm obs})
=
\sum_{j}
\tau_{\rm eff}^{j}(z_j),
\label{eq:tau_lyman}
\end{equation}
where $\tau_{\rm eff}^j(z_j)$ is the effective opacity of transition $j$ at redshift $z_j$ such that $(1+z_j)\lambda_j=\lambda_{\rm obs}$, and $\lambda_j$ is the wavelength of transition $j$ in the rest frame. To implement this, we utilized Sherwood and Sherwood-Relics simulations \citep{Bolton2017, Puchwein2023} to determine the effective optical depth for each LyS line across a range of absorption redshifts and $\Gamma_{\rm HI}$ values.
The simulation snapshots are separated by redshift intervals of $\Delta z = 0.1$, and each snapshot contains 5000 simulated lines of sight. At each redshift, the native \lya\ optical depths are rescaled so that the simulated mean \lya\ transmission matches observational measurements. The optical depths of higher-order LyS transitions are then obtained by scaling the \lya\ optical depth according to $\tau_j/\tau_\alpha = (f_j\lambda_j)/(f_\alpha\lambda_\alpha)$, where $f_j$ and $\lambda_j$ are the oscillator strength and rest-frame wavelength of the $j$th transition.
The mean \lya\ transmission is related to the effective optical depth through $\langle T_{\mathrm{Ly}\alpha}\rangle = e^{-\tau_{\mathrm{eff}}^\alpha}$. We adopt the measurements of ~\citet{Becker2013} for $z \leq 4$ and those of~\citet{Bosman2022} for $z \geq 5$. Over the intermediate redshift range, $4.0 < z < 5$, we follow the prescription of~\citet{Boera2019} and interpolate using the power-law relation $\tau_{\mathrm{eff}}^\alpha(z)=1.56[(1+z)/5.75]^{4.0}$.

Throughout this work, we include all hydrogen LyS transitions from $n=2$ to $n=1000$. The rest-frame wavelengths were computed using the Rydberg formula, 
$$
\lambda_n^{-1}
=
R_{\rm H}
\left(
1-\frac{1}{n^{2}}
\right),
$$
where $R_{\rm H}=1.0967758\times10^{7}\,\mathrm{m^{-1}}$ is the Rydberg constant for hydrogen.
The oscillator strengths for the $1s \rightarrow np$ transitions were evaluated using the closed-form analytic hydrogenic expression
\begin{equation}
f_{1n} = \frac{256\, n^{5}(n-1)^{2n-4}}{3\,(n+1)^{2n+4}},
\end{equation}
originally derived by \citet{Menzel1935} (see also \citealt{Bethe1957}, \citealt{Cowley2000O} and \citealt{Rohrmann2022}). We evaluated $f_{1n}$ in log-space to avoid numerical overflow at large $n$, extending the tabulation to $n=1000$. 
The truncation at $n=1000$ is conservative. The oscillator strengths of
high-order Lyman transitions decrease rapidly, with $f_{1n}\propto n^{-3}$
at large $n$, so the contribution of increasingly high-order transitions
becomes progressively smaller. We explicitly tested the convergence of the
full transmission calculation and found no noticeable change in the predicted
transmission when including transitions beyond $n=100$. We therefore adopt
$n_{\rm max}=1000$ throughout this work.

Because the wavelengths of very high-order LyS transitions converge rapidly toward the Lyman limit, the spacing between adjacent transitions becomes increasingly small. To reduce the computational cost of the transmission calculations, transitions with rest-frame wavelengths shorter than 912.7~\AA\ were grouped into wavelength bins of width $\Delta\lambda=0.1$ ~\AA. Within each bin, an effective line was constructed by assigning it a representative wavelength equal to the mean wavelength of all transitions contained in the bin, and an oscillator strength equal to the sum of the oscillator strengths of the constituent transitions. This procedure preserves the total oscillator strength within each wavelength bin while substantially reducing the computational expense of cumulative higher-order LyS transmission.

The next step is to account for LyC absorption at rest-frame wavelengths below the Lyman limit.
The effective LyC optical depth for a photon emitted by a quasar at redshift $z_{\rm qso}$ and observed at the Lyman limit ($912$~\AA) at redshift $z_{\rm LL}$ is given by

\begin{equation}
\tau_{\rm eff}^{\rm LyC}(z_{\rm LL},z_{\rm qso})
=
\frac{c}{H_0 \Omega_m^{1/2}}
(1+z_{\rm LL})^{2.75}
\int_{z_{\rm LL}}^{z_{\rm qso}}
\kappa_{\rm LL}(z')
(1+z')^{-5.25}
\,dz' ,
\end{equation}
 The wavelength
dependence of the ionizing absorption cross-section is approximated
as $
\sigma(\lambda) \propto \lambda^{2.75},
$
following \citet{OMeara2013} and \citet{Worseck2014}.

For rest-frame wavelengths shorter than $912$~\AA, both LyC and foreground LyS absorption contribute to the total effective optical depth. The total effective optical depth is therefore given by
\begin{equation}
\tau_{\rm eff}^{\rm tot}(\lambda_{\rm obs})
=
\tau_{\rm eff}^{\rm LyC}(\lambda_{\rm obs})
+
\sum_j
\tau_{\rm eff}^{j}(z_j),
\end{equation}
where the summation extends over all foreground LyS transitions, from \lya\ to the highest-order transition included in the calculation. For rest-frame wavelengths longer than $912$~\AA, the LyC term is absent and only foreground LyS absorption contributes to the total effective optical depth.

One complication is the quasar proximity effect, which can significantly modify the transmission near the quasar (see Fig.~\ref{fig:illustration} for an illustration). The enhanced ionizing radiation from the quasar reduces the neutral hydrogen fraction and, consequently, the opacity of the IGM within the proximity zone. This effect must therefore be incorporated into the transmission calculations.

To account for this effect, we follow the methodology of \citet{Becker2021}, who extended the \mfp\ measurement methodology of \citet{Prochaska2009} to explicitly include the effect of the quasar proximity zone. Motivated by numerical simulations, the ionizing opacity near the quasar is assumed to depend on the local $\Gamma_{\rm HI}$, such that regions exposed to stronger ionizing radiation become more transparent. Specifically, the LyC opacity, $\kappa_{\rm LL}$, is parameterized as a power law of the photoionization rate,

\begin{equation}\label{eq:kappa}
    \kappa_{\rm LL} (r) = \kappa_{\rm LL}^{\rm bg} \left[  1+\frac{\Gamma_{\rm qso}(r)}{\Gamma_{\rm bg}} \right]^{-\xi} ~,
\end{equation}
where $\Gamma_{\rm qso}(r)$ is the photoionization rate due to the QSO at a distance $r$ from the QSO, and $\Gamma_{\rm bg}$ is the background photoionization rate. We assume $\xi=0.67$; the dependence of our results on this parameter is discussed in Section~\ref{subsec:Sensitivity_to_xi}.

We assume $\kappa_{\rm LL}^{\rm bg} = 1/\lambda_{\rm mfp}$, which is appropriate for the short mean free path values at these redshifts. When constructing our transmission models we calculate \mfp\ for each simulation directly, rather than applying measured values.  The goal here is to determine whether the models self-consistently reproduce both the observed LyS and LyC transmission.
To calculate \mfp, the simulated neutral hydrogen fractions are first rescaled so that the simulated mean \lya\ transmission matches the observed value. The neutral hydrogen column density of every simulation pixel is then computed from the local gas density, neutral fraction, and pixel size. Long sightlines are constructed by concatenating multiple randomly selected and randomly shifted simulation skewers, and synthetic LyC transmission spectra are generated by integrating the cumulative photoionization opacity along each sightline. The optical depth contributed by each simulation pixel is computed as
$
\tau_{\rm LyC}(\lambda)
=
N_{\rm HI}\,
\sigma_{\rm LL}
\left(
\frac{\lambda_{\rm rest}}
{\lambda_{\rm LL}}
\right)^{2.75}.
$
where $N_{\rm HI}$ is the neutral hydrogen column density of the pixel, $\sigma_{\rm LL}$ is the hydrogen photoionization cross section at the Lyman limit, and $\lambda_{\rm rest}$ is the photon wavelength in the rest frame of the absorber. The cumulative optical depth is obtained by summing the contributions from all intervening pixels, and the corresponding transmission is given by $T(\lambda)=\exp[-\tau_{\rm LyC}(\lambda)]$. 
The mean LyC transmission profile is computed by averaging over 2000 independent random sightlines. The simulation \mfp\ is then defined as the proper distance from the source redshift to the point where the mean LyC transmission decreases to $e^{-1}$. We verified that the inferred \mfp\ is well converged for this number of realizations.

In order to calculate $\Gamma_{\rm qso}(r)$, a key parameter used to describe the proximity zone effect in \cite{Becker2021} is $R_{\rm eq}$. It denotes the distance from the QSO where $\Gamma_{\rm qso}(r)$ and $\Gamma_{\rm bg}$ would be equal for purely geometric dilution in the absence of attenuation.
 $R_{\rm eq}$ is related to $\Gamma_{\rm bg}$ and the ionizing luminosity of the QSO, $L_{912}$, by

\begin{equation}\label{eq:req}
    R_{\rm eq} = \left[ \frac{L_{912} \sigma_0}{4\pi h \Gamma_{\rm bg}(\alpha_{\nu}^{\rm ion} + 2.75)}  \right]^{1/2} ~,
\end{equation}
where $\sigma_0$ is the \hi\ photoionization cross section at 912~\AA, $\alpha_{\nu}^{\rm ion}$ is the power-law index of the quasar continuum at $\lambda < 912$~\AA\ in the frequency domain, and $h$ is the  Planck constant \citep{Calverley2011}.
$L_{912}$ can be further related to the absolute magnitude corresponding to the mean luminosity at rest-frame 1450~\AA, $L_{1450}$, through
\begin{equation}
    L_{912}=L_{1450}\left(\frac{\nu_{912}}{\nu_{1450}}\right)^{-\alpha_{\nu}^{\rm UV}}.
\end{equation}
Here, $\nu_{912}$ and $\nu_{1450}$ are the photon frequencies corresponding to 912~\AA\ and 1450~\AA, respectively, and $\alpha_{\nu}^{\rm UV}$ is the power-law slope of the non-ionizing quasar continuum (912~\AA $< \lambda <$ 1450~\AA).

The $R_{\rm eq}$ values for the individual quasars in our sample were originally calculated by \citet{Zhu2023} assuming the $\Gamma_{\rm bg}$ values measured by \citet{Gaikwad2023}. Since each of our simulations predicts a different value of $\Gamma_{\rm bg}$, we rescale the $R_{\rm eq}$ of each quasar using Equation~\ref{eq:req} with the corresponding simulation-specific $\Gamma_{\rm bg}$. The rescaled values are then averaged over the quasars contributing to each composite, and the mean value is applied to the respective simulated transmission model at that redshift. The resulting mean $R_{\rm eq}$ values are listed in Table~\ref{tab:req_composites}.

\begin{table}
\centering
\caption{Mean $R_{\rm eq}$ values for the four quasar composites. All values are given in proper Mpc (pMpc).}
\label{tab:req_composites}
\begin{tabular}{c|ccccc}
\hline
$z_{\rm comp}$ & Sherwood & Relics53 & Relics57 & Relics60 & Relics66 \\
\hline
5.93 & 9.96 & 14.66 & 14.84 & 15.57 & 14.07 \\
5.64 & 10.30 & 14.65 & 14.85 & 15.02 & 13.69 \\
5.32 & 7.03 & 9.64 & 9.45 & 9.47 & 8.78 \\
5.08 & 4.63 & 6.13 & 5.96 & 5.98 & 5.61 \\
\hline
\end{tabular}
\end{table}

We then include the proximity zone effect for LyS and LyC by matching the effective optical depth to a $\Gamma_{\rm HI}$ value that corresponds to $\Gamma_{\rm bg} + \Gamma_{\rm qso}$ as a function of distance from the QSO. 

We compute $\Gamma_{\rm qso}$ by dividing the line of sight into small steps of distance $\delta r$, and solving for $\Gamma_{\rm qso} (r)$ numerically using the method described in \cite{Becker2021}. For the first step we assume that $\Gamma_{\rm qso}$ decreases purely geometrically, i.e.
\begin{equation}\label{eq:Gamma_qso_0}
    \Gamma_{\rm qso}(r=\delta r) = \Gamma_{\rm bg} \left( \frac{r}{R_{\rm eq}} \right)^{-2} \, .
\end{equation}
We then solve for $\Gamma_{\rm qso}(r + \delta r)$ over subsequent steps as 
\begin{equation}\label{eq:Gamma_qso_r}
    \Gamma_{\rm qso}(r + \delta r) = \Gamma_{\rm qso}(r) \left( \frac{r + \delta r}{r} \right)^{-2} e^{-\kappa_{\rm LL}(r) \delta r} \, ,
\end{equation}
where $\kappa_{\rm LL}(r)$ is computed using Equation~(\ref{eq:kappa}).
Finally, $\tau_{\rm eff}^{\rm LyC}(\lambda_{\rm obs})$, $\tau_{\rm eff}^j(z_j)$ and $\tau_{\rm eff}^{\rm Lyman} (\lambda_{\rm obs})$ can be computed.

We note that our fiducial calculation of the Lymnan-series transmission implicitly neglects correlations among the individual LyS transitions and LyC absorption. In Appendix~\ref{app:correlations}, we test the impact of this approximation using a sightline-based calculation and find that these correlations have only a small effect on the predicted mean transmission.

\subsection{Treatment of neutral regions inside quasar proximity zones}
\label{Treatment-of-Neutral-Regions}

For the Sherwood simulation, the rescaling procedure described in Section~\ref{subsec:methods} is straightforward, because the gas is assumed to remain in photoionization equilibrium, implying $\tau \propto \Gamma^{-1}$. In contrast, the original Sherwood--Relics simulations contain neutral regions with $\Gamma_{\rm HI}=0$.
The IGM around bright quasars is typically expected to reionize earlier than the rest of the IGM, however.  The simulation boxes we use are too small to contain rare, bright quasars, which prompts us to employ a simple model for including the ionized proximity zones. Specifically, we assume that the gas around quasars is ionized and in photoionization equilibrium out to a distance where the ionization rate from the quasar would equal the background photoionization rate. We therefore construct a companion set of equilibrium realizations for each of the Sherwood--Relics simulations, generated using a spatially uniform ionizing background while keeping all other parameters fixed, to determine the new equilibrium neutral fractions and corresponding optical depths. Inside the ionized region, the companion equilibrium simulation is adopted, while outside it the original Sherwood--Relics realization is retained. The corresponding redshift extents of ionized regions are $\Delta z=[0.12,0.13,0.10,0.06]$ for the Relics53 simulation, ordered by the composite redshifts $\langle z \rangle=[5.93,5.64,5.32,5.08]$. Using this combined simulation, we then apply our method to determine the transmission.

This procedure has a negligible effect on composites whose average redshift lies below the reionization completion redshift of the adopted simulation, since the IGM is already in photoionization equilibrium. For composites whose average redshift exceeds the reionization completion redshift, however, replacing the non-equilibrium region with the equilibrium realization increases the transmission in the vicinity of the quasar. In Appendix~\ref{sec:Sensitivity_to_the_Definition_of_the Quasar-affected Region}, we illustrate the effect of this treatment on the different quasar composites and show that the results are not strongly sensitive to the choice of the transition distance.

\subsection{Composite mean transmission}
\label{subsec:Composite_Mean_Transmission}
For each simulation, we compute the expected mean transmission profile for each composite given the mean redshift and simulation-dependent mean $R_{\rm eq}$ value of the contributing quasars. The resulting transmission profiles over the rest-frame wavelength range 820--1220~\AA\ are shown in Fig.~\ref{fig:transmissions_zeta67}. Unless otherwise stated, all results presented here assume $\xi = 0.67$, which we adopt as our fiducial value. The dependence of the transmission profiles on this parameter is discussed in Section~\ref{subsec:Sensitivity_to_xi}.

An alternative approach is to compute a mean transmission profile for each individual quasar using its own redshift and simulation-dependent \Req, and then average the resulting transmission profiles within each composite. As shown in Appendix~\ref{sec:Alternative_Transmission_Calculations}, we verify that this procedure produces transmission profiles that are nearly identical over the wavelength range of interest. We therefore adopt the average redshift and \Req\ approach throughout this work because it is computationally more efficient.

\subsection{Assessing the agreement between data and models}

To quantitatively compare the simulated mean transmission profiles with the observed composites, we compute the  $\chi^2$ statistic,

\begin{equation}
\chi^2 =
(\mathbf{D}-\mathbf{M})^{\rm T}
\mathbf{C}^{-1}
(\mathbf{D}-\mathbf{M}),
\end{equation}
where $\mathbf{D}$ and $\mathbf{M}$ are the observed and simulated composite transmission vectors, respectively, and $\mathbf{C}$ is the covariance matrix. The reduced $\chi^2$ is defined as $\chi^2_\nu = \chi^2/N_{\rm dof}$, where $N_{\rm dof}$ is the number of degrees of freedom, equal to the number of wavelength bins included in the comparison. For each composite, we evaluate $\chi^2$ up to 950~\AA, with lower limits of $830$, $830$, $850$, and $860$~\AA\ at $z=5.08$, $5.32$, $5.64$, and $5.93$, respectively, where the predicted transmission in most models approaches zero and/or the observed spectra become increasingly noisy.  These ranges contain $337$, $337$, $278$, and $249$ wavelength bins of $120~{\rm km~s^{-1}}$, respectively. The upper limit of $950$~\AA\ approximately corresponds to Ly$\delta$, keeping the comparison focused on the higher-order LyS region and away from nearby strong emission features. We also perform a separate $\chi^2$ comparison over 912--950~\AA\ to isolate the contribution of LyS absorption from LyC absorption.

The observational covariance matrix cannot be estimated reliably from the data alone because each composite contains too few quasars. Instead, we estimate the covariance from Monte Carlo realizations of the simulations. For each simulation, we construct a mock composite by generating a long simulated line of sight for each quasar, using its redshift and \Req. We then match the instrumental resolution of each observed spectrum by convolving the simulated spectrum with a Gaussian kernel with the corresponding FWHM, accounting for changes in resolution across the spectrum where necessary. The simulated spectra are then binned onto the same wavelength grid as the corresponding observed spectra, and observational Gaussian noise is added using their respective noise arrays. Finally, the mock spectra are combined following the same procedure used to construct the observed composite. This process is repeated 5000 times, producing an ensemble of mock composites.

The covariance matrix is then estimated from the ensemble of mock composites as
\begin{equation}
C_{ij}=
\frac{1}{N-1}
\sum_{k=1}^{N}
\left(
T_{k,i}-\bar{T}_i
\right)
\left(
T_{k,j}-\bar{T}_j
\right),
\end{equation}
where $T_{k,i}$ is the transmission in wavelength bin $i$ for the $k$th mock realization, $\bar{T}_i$ is the mean transmission over all realizations, and $N=5000$ is the number of mock composites. This approach naturally incorporates both the intrinsic sightline-to-sightline variance predicted by the simulations and the pixel-to-pixel noise in the observations.

The covariance matrix estimated from the mock realizations is subsequently rescaled to match the variance implied by the observational uncertainties. Specifically, we compute the ratio of the diagonal elements of the Monte Carlo covariance matrix to the squared observational uncertainties in each wavelength bin and use the median of these ratios as a single normalization factor. The entire covariance matrix is then rescaled by this factor, preserving the correlation structure predicted by the simulations while ensuring that its diagonal is consistent with the variance implied by the observational uncertainties. The resulting rescaled covariance matrix is adopted in the $\chi^2$ analysis. For all model comparisons, we adopt the covariance matrix derived from the Relics53 mock realizations, as this model provides the best overall agreement with the observed transmission. We have verified, however, that using covariance matrices derived from any of the other simulations does not affect the relative ranking of the models or our conclusions.

For our fiducial value of $\xi=0.67$, the resulting $\chi^2$ and $\chi^2_\nu$ values for each composite and simulation are presented in Table~\ref{tab:chi2_xi_comparison}. The Sherwood simulation provides the poorest agreement with the observations, with a combined $\chi^2=2352.0$ ($\chi^2_\nu=1.96$). The Sherwood--Relics models yield substantially lower values, with combined $\Delta\chi^2\equiv\chi^2_{\rm Sherwood}-\chi^2_{\rm Relics}=517$--$870$. Among these, the Relics53 model provides the best overall match, with a combined $\chi^2=1481.7$ ($\chi^2_\nu=1.23$). All four Sherwood--Relics models yield lower $\chi^2$ values than Sherwood in every redshift bin, with the largest improvement occurring at $z=5.93$. At this redshift, $\chi^2$ decreases from $649.0$ for Sherwood to $353.3$ for Relics53, compared to a decrease from $415.7$ to $266.2$ at $z=5.08$. Among the Sherwood--Relics models, the agreement generally worsens with increasing reionization completion redshift, with the largest differences between these models occurring at $z=5.32$. The Relics53 model yields the lowest $\chi^2$ in all four redshift bins. Although the Sherwood--Relics models provide substantially better agreement with the observations, none yields a combined $\chi^2_\nu$ of unity. For Relics53, the individual values range from $\chi^2_\nu=0.79$ at $z=5.08$ to $1.54$ at $z=5.64$. These deviations from $\chi^2_\nu = 1$, even for well-fitting models, may occur due to limitations in the estimated covariance matrix. We therefore emphasize the relative differences in $\chi^2$ between simulations rather than relying solely on their absolute goodness of fit.

To assess whether the preference for the Sherwood--Relics models is driven primarily by LyC absorption, we repeat the $\chi^2$ analysis over the rest-frame wavelength interval 912--950~\AA, where only LyS absorption contributes. The results are presented in Table~\ref{tab:chi2_lys_only}. For our fiducial value of $\xi=0.67$, Sherwood again provides the poorest overall agreement with the observations, with a combined $\chi^2=981.7$ ($\chi^2_\nu=2.41$). The Sherwood--Relics models yield substantially lower values, with combined $\Delta\chi^2=271$--$400$. Among these, Relics53 provides the best overall match, with a combined $\chi^2=581.3$ ($\chi^2_\nu=1.42$). The largest improvement occurs at $z=5.93$, where $\chi^2$ decreases from $414.7$ for Sherwood to $213.1$ for Relics53. In contrast, the differences are much smaller at $z=5.08$, where Relics66 provides the lowest $\chi^2$ among the models considered. These results demonstrate that the better agreement of the Sherwood--Relics models with the observations is not driven solely by differences in LyC absorption, but is also evident in the higher-order LyS transmission.

\begin{figure}
    \centering
    \includegraphics[width=\columnwidth]{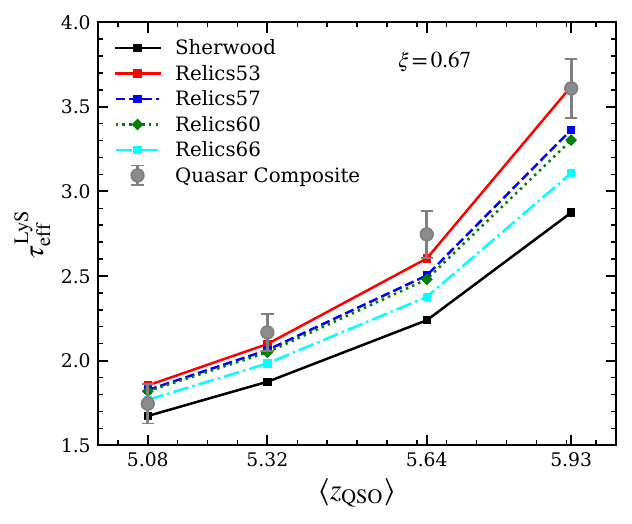}
    \caption{
    Evolution of the effective higher-order LyS
    optical depth, $\tau_{\rm eff}^{\rm LyS}$, measured
    over the rest-frame wavelength interval 912--950~\AA.
    Gray circles show the observed quasar composites,
    while colored lines show predictions from the Sherwood
    and Sherwood--Relics simulations for the fiducial
    value of $\xi=0.67$. 
    }
    \label{fig:tau_eff_lys}
\end{figure}

To further characterize the evolution of higher-order LyS
absorption, we measure the effective optical depth of each
observed quasar composite over the rest-frame wavelength
interval 912--950~\AA. We calculate
$\tau_{\rm eff}^{\rm LyS}=-\ln\langle T\rangle$, where
$\langle T\rangle$ is the mean transmission over this
wavelength interval. The uncertainties are estimated using
10,000 bootstrap realizations of each observed composite.
We also include an assumed 10\% uncertainty in the intrinsic
continuum normalization by multiplying the continuum in
each bootstrap realization by a random factor drawn from
a Gaussian distribution with a mean of unity and a standard
deviation of 0.1. The final uncertainties
are determined from the standard deviation of the resulting
$\tau_{\rm eff}^{\rm LyS}$ distribution.

The resulting measurements are shown in
Fig.~\ref{fig:tau_eff_lys} alongside predictions from
the Sherwood and Sherwood--Relics simulations for our
fiducial value of $\xi=0.67$. We measure
$\tau_{\rm eff}^{\rm LyS}=1.745\pm0.117$,
$2.167\pm0.110$, $2.746\pm0.139$, and
$3.608\pm0.174$ at $z=5.08$, $5.32$, $5.64$,
and $5.93$, respectively. The observed effective
optical depth increases substantially with redshift,
indicating stronger cumulative LyS absorption toward
higher redshifts. Sherwood systematically predicts lower
effective optical depths than observed, with the differences
becoming more pronounced at higher redshifts. In contrast,
the Sherwood--Relics models generally provide better
agreement with the observed evolution, with Relics53
providing the best overall match, consistent with
the results of the wavelength-dependent $\chi^2$ analysis. The corresponding results for $\xi=0.33$ and $\xi=1.0$
are presented in Fig.~\ref{fig:tau_eff_lys_appendix}
in Appendix~\ref{sec:Transmission_Profiles_for_Alternative_Values}.

\begin{table*}
\centering

\caption{$\chi^2$ values for the four quasar composites compared with the five simulation models, with $\chi^2/N_{\rm pix}$ given in parentheses. The wavelength intervals are 860--950, 850--950, 830--950, and 830--950~\AA\ for the composites at $z=5.93$, 5.64, 5.32, and 5.08, respectively. Results are shown for three values of $\xi$. For each $\xi$, the final row gives the combined values for all four composites.}
\label{tab:chi2_xi_comparison}
\begin{tabular}{ccccccc}
\hline
$\xi$ & Composite & Sherwood & Relics53 & Relics57 & Relics60 & Relics66 \\
\hline

\multirow{5}{*}{$0.33$} & $z=5.93$ & 517.1 (2.08) & 355.5 (1.43) & 355.6 (1.43) & 357.9 (1.44) & 378.6 (1.52) \\
 & $z=5.64$ & 518.8 (1.87) & 426.5 (1.53) & 432.4 (1.56) & 434.0 (1.56) & 450.4 (1.62) \\
 & $z=5.32$ & 661.8 (1.96) & 435.2 (1.29) & 443.0 (1.31) & 452.4 (1.34) & 519.6 (1.54) \\
 & $z=5.08$ & 380.3 (1.13) & 263.8 (0.78) & 265.1 (0.79) & 282.4 (0.84) & 330.8 (0.98) \\
& \textbf{Combined} & \textbf{2078.0 (1.73)} & \textbf{1480.9 (1.23)} & \textbf{1496.0 (1.25)} & \textbf{1526.6 (1.27)} & \textbf{1679.4 (1.40)} \\
\hline
\multirow{5}{*}{$0.67$} & $z=5.93$ & 649.0 (2.61) & 353.3 (1.42) & 364.1 (1.46) & 371.8 (1.49) & 427.8 (1.72) \\
 & $z=5.64$ & 566.9 (2.04) & 428.0 (1.54) & 442.7 (1.59) & 447.2 (1.61) & 479.4 (1.72) \\
 & $z=5.32$ & 720.3 (2.14) & 434.2 (1.29) & 450.6 (1.34) & 468.8 (1.39) & 564.7 (1.68) \\
 & $z=5.08$ & 415.7 (1.23) & 266.2 (0.79) & 274.9 (0.82) & 298.3 (0.89) & 362.7 (1.08) \\
& \textbf{Combined} & \textbf{2352.0 (1.96)} & \textbf{1481.7 (1.23)} & \textbf{1532.3 (1.28)} & \textbf{1586.1 (1.32)} & \textbf{1834.7 (1.53)} \\
\hline
\multirow{5}{*}{$1.0$} & $z=5.93$ & 802.8 (3.22) & 355.2 (1.43) & 376.6 (1.51) & 394.7 (1.58) & 503.0 (2.02) \\
 & $z=5.64$ & 618.2 (2.22) & 434.6 (1.56) & 454.3 (1.63) & 462.9 (1.67) & 516.1 (1.86) \\
 & $z=5.32$ & 774.1 (2.30) & 441.2 (1.31) & 464.8 (1.38) & 491.3 (1.46) & 612.1 (1.82) \\
 & $z=5.08$ & 445.2 (1.32) & 272.8 (0.81) & 287.1 (0.85) & 315.8 (0.94) & 391.6 (1.16) \\
& \textbf{Combined} & \textbf{2640.3 (2.20)} & \textbf{1503.8 (1.25)} & \textbf{1582.8 (1.32)} & \textbf{1664.7 (1.39)} & \textbf{2022.8 (1.68)} \\
\hline
\end{tabular}
\end{table*}

\begin{table*}
\centering
\caption{
Same as Table~\ref{tab:chi2_xi_comparison}, but restricted
to the rest-frame wavelength interval 912--950~\AA,
where the absorption arises entirely from LyS
transitions.
}
\label{tab:chi2_lys_only}
\begin{tabular}{ccccccc}
\hline
$\xi$ & Composite & Sherwood & Relics53 & Relics57 & Relics60 & Relics66 \\
\hline
\multirow{5}{*}{$0.33$} & $z=5.93$ & 383.4 (3.76) & 218.0 (2.14) & 216.7 (2.12) & 220.7 (2.16) & 248.8 (2.44) \\
 & $z=5.64$ & 254.0 (2.49) & 160.3 (1.57) & 168.7 (1.65) & 170.9 (1.68) & 191.4 (1.88) \\
 & $z=5.32$ & 216.4 (2.12) & 123.6 (1.21) & 128.2 (1.26) & 130.3 (1.28) & 149.8 (1.47) \\
 & $z=5.08$ & 77.7 (0.76) & 80.2 (0.79) & 72.7 (0.71) & 71.9 (0.70) & 64.9 (0.64) \\
& \textbf{Combined} & \textbf{931.6 (2.28)} & \textbf{582.1 (1.43)} & \textbf{586.4 (1.44)} & \textbf{593.8 (1.46)} & \textbf{654.8 (1.60)} \\
\hline
\multirow{5}{*}{$0.67$} & $z=5.93$ & 414.7 (4.07) & 213.1 (2.09) & 226.2 (2.22) & 234.4 (2.30) & 282.6 (2.77) \\
 & $z=5.64$ & 265.9 (2.61) & 165.0 (1.62) & 177.2 (1.74) & 181.0 (1.77) & 206.7 (2.03) \\
 & $z=5.32$ & 222.4 (2.18) & 125.3 (1.23) & 131.3 (1.29) & 134.6 (1.32) & 156.7 (1.54) \\
 & $z=5.08$ & 78.7 (0.77) & 77.9 (0.76) & 71.1 (0.70) & 70.0 (0.69) & 64.5 (0.63) \\
& \textbf{Combined} & \textbf{981.7 (2.41)} & \textbf{581.3 (1.42)} & \textbf{605.8 (1.48)} & \textbf{620.0 (1.52)} & \textbf{710.5 (1.74)} \\
\hline
\multirow{5}{*}{$1.0$} & $z=5.93$ & 435.6 (4.27) & 214.8 (2.11) & 235.9 (2.31) & 248.3 (2.43) & 308.6 (3.03) \\
 & $z=5.64$ & 273.5 (2.68) & 169.9 (1.67) & 183.8 (1.80) & 189.3 (1.86) & 218.1 (2.14) \\
 & $z=5.32$ & 226.0 (2.22) & 126.7 (1.24) & 133.5 (1.31) & 137.6 (1.35) & 161.7 (1.59) \\
 & $z=5.08$ & 79.4 (0.78) & 76.6 (0.75) & 70.2 (0.69) & 69.0 (0.68) & 64.3 (0.63) \\
& \textbf{Combined} & \textbf{1014.5 (2.49)} & \textbf{588.0 (1.44)} & \textbf{623.4 (1.53)} & \textbf{644.3 (1.58)} & \textbf{752.8 (1.85)} \\
\hline
\end{tabular}
\end{table*}

\subsection{Sensitivity to $\xi$}
\label{subsec:Sensitivity_to_xi}

Previous analyses of direct mean-free-path measurements have generally adopted a fiducial value of $\xi=0.67$, motivated by theoretical models of ionizing sinks \citep{Miralda-Escud2000,Furlanetto2005,McQuinn2011}, while recognizing that its true value remains uncertain (see \citealt{Becker2021}). More recently, \citet{Feron2024} investigated the dependence of the local Lyman-limit attenuation length, $\lambda=(n_{\rm HI}\sigma_{912})^{-1}$, on the photoionization rate, $\lambda\propto\Gamma_{\rm HI}^{\xi}$, in the ionized regions of the Relics57 simulation. They found a value of $\xi\sim0.8$ at $z\sim6.3$ (their Fig.~C2), although they caution against extrapolating these results to lower redshifts.
To account for this uncertainty, \citet{Becker2021} and \citet{Zhu2023} explored values in the range $0.3 \leq \xi \leq 1.0$ when inferring \mfp\,. In this work, we likewise adopt the fiducial value $\xi=0.67$ and investigate how varying $\xi$ affects the predicted LyS transmission profiles. The transmission spectra obtained for the fiducial value are presented in Fig.~\ref{fig:transmissions_zeta67}, while the corresponding predictions for $\xi=0.33$ and $\xi=1.0$ are shown in Figs.~\ref{fig:transmissions_zeta33} and ~\ref{fig:transmissions_zeta100}, respectively. As expected from Equation~\ref{eq:kappa}, increasing $\xi$ increases the magnitude of the proximity effect, resulting in systematically higher LyS and LyC transmission, while decreasing $\xi$ has the opposite effect. The corresponding  $\chi^2$ and $\chi^2_\nu$ values are summarized in Tables~\ref{tab:chi2_xi_comparison} and \ref{tab:chi2_lys_only} for the full and LyS-only wavelength intervals, respectively.

For $\xi=0.33$, the Sherwood simulation yields a combined $\chi^2=2078.0$ ($\chi^2_\nu=1.73$), while the Sherwood--Relics models provide substantially better agreement with the observations. The lowest combined $\chi^2$ is obtained for Relics53, with $\chi^2=1480.9$ ($\chi^2_\nu=1.23$), followed by Relics57 ($\chi^2=1496.0$). For $\xi=1.0$, the combined $\chi^2$ for Sherwood increases to $2640.3$ ($\chi^2_\nu=2.20$), whereas Relics53 continues to provide the best overall match, with $\chi^2=1503.8$ ($\chi^2_\nu=1.25$).

The dependence on $\xi$ is considerably stronger for Sherwood than for Relics53, particularly at the highest redshift. At $z=5.93$, the Sherwood $\chi^2$ increases from $517.1$ to $802.8$ as $\xi$ increases from $0.33$ to $1.0$, while the corresponding values for Relics53 remain nearly unchanged. The combined $\chi^2$ for Relics53 also varies only modestly across the three values of $\xi$, ranging from $1480.9$ to $1503.8$. A similar trend is found in the LyS-only comparison (Table~\ref{tab:chi2_lys_only}), where Relics53 provides the lowest combined $\chi^2$ for all three values of $\xi$, ranging from $581.3$ to $588.0$, compared to $931.6$--$1014.5$ for Sherwood. Thus, our main conclusion that the late-reionization models provide better agreement with the observed LyS transmission than Sherwood remains unchanged.

\begin{figure*}
    \centering
    \includegraphics[width=\linewidth]{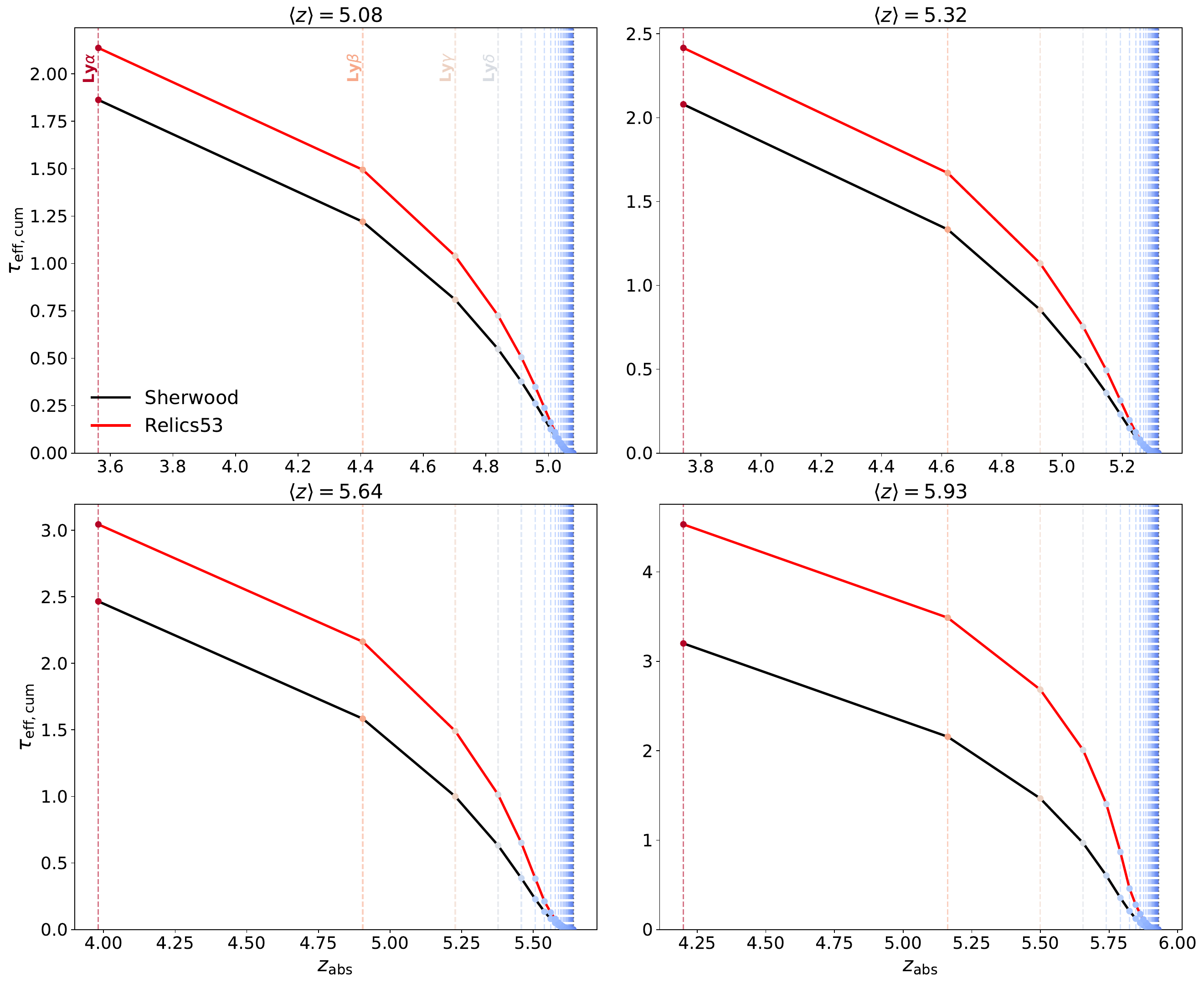}
    \caption{
Cumulative LyS effective optical depth at $\lambda_{\rm rest}=912$~\AA\ for Sherwood (black) and Relics53 (red) for the four quasar composites. The cumulative optical depth is constructed by successively adding the contribution of each LyS transition, beginning with the highest absorber redshift, closest to the quasar, and proceeding toward lower redshift. The colored points and vertical dashed lines mark the absorber redshift associated with each individual LyS transition, with selected transitions labeled in the upper-left panel.
}
    \label{fig:cumulative_tau_eff_912}
\end{figure*}

\begin{figure*}
    \centering
    \includegraphics[width=\linewidth]{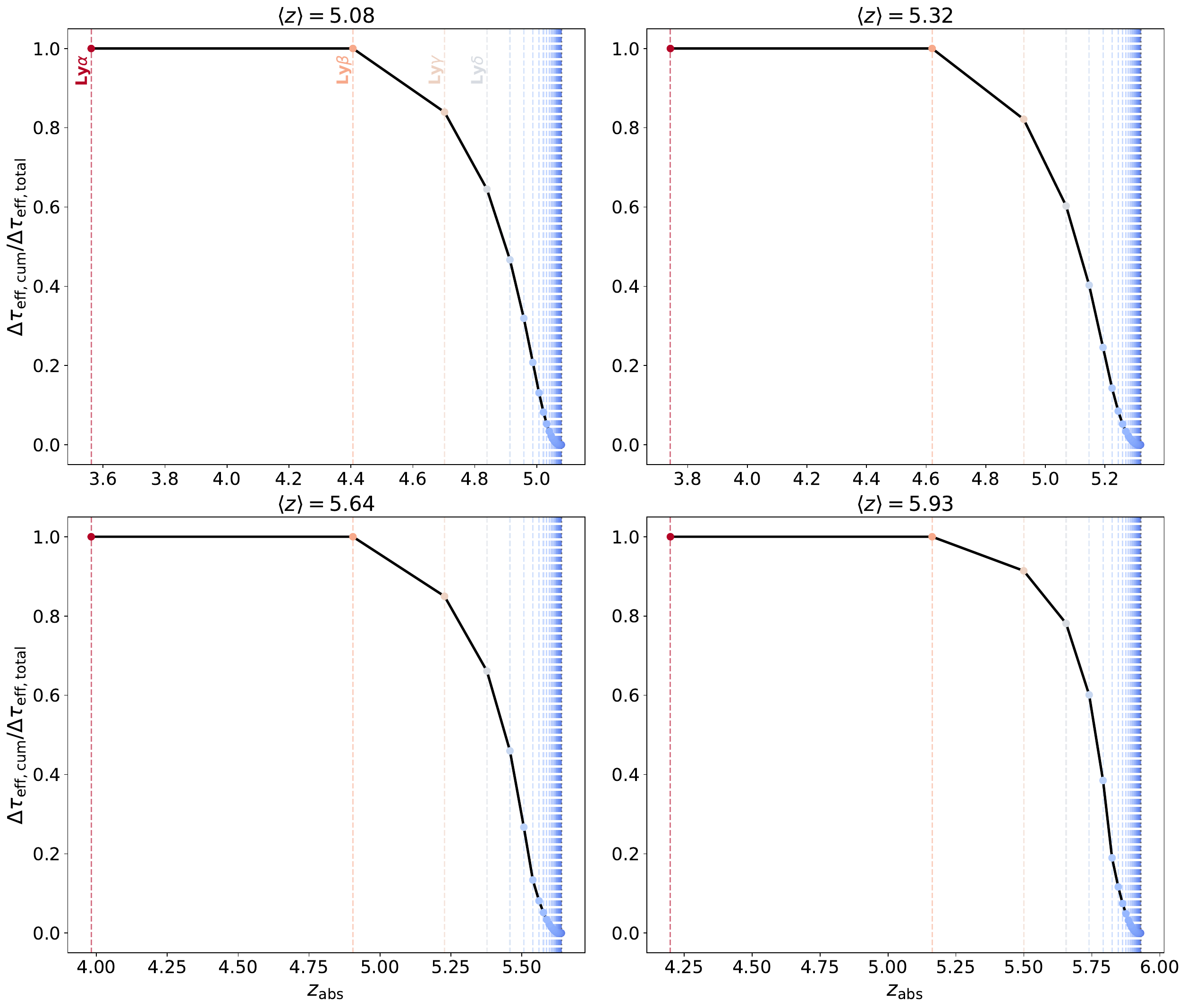}
    \caption{ 
Cumulative fraction of the total difference in LyS effective optical depth between Relics53 and Sherwood at $\lambda_{\rm rest}=912$~\AA\ for the four quasar composites. The cumulative $\Delta\tau_{\rm eff}$ is constructed by successively adding the contribution of each LyS transition, beginning at the highest absorber redshift, closest to the quasar, and proceeding toward lower redshift. The resulting cumulative difference is normalized by the total difference from all LyS transitions. The colored points and vertical dashed lines mark the absorber redshift associated with each individual LyS transition, with selected transitions labeled in the upper-left panel.
}
    \label{fig:cumulative_delta_tau_eff_912}
\end{figure*}

\begin{figure*}
    \centering
    \includegraphics[width=\linewidth]{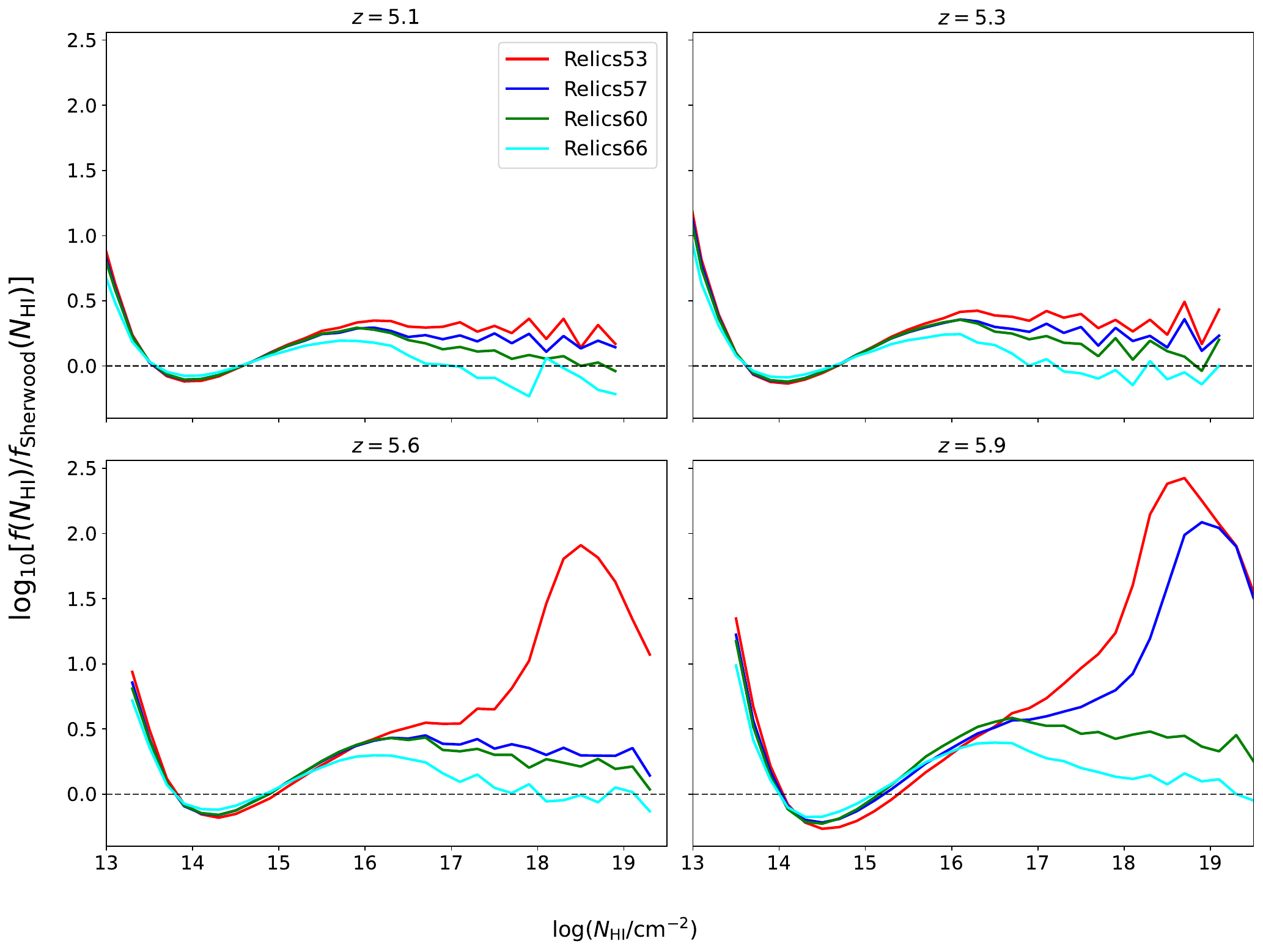}
\caption{Ratio of $f(N_{\rm HI})$ predicted by the Relics53, Relics57, Relics60, and Relics66 simulations relative to the Sherwood simulation at $z=5.1$, 5.3, 5.6, and 5.9.}
    \label{fig:cddf}
\end{figure*}

\section{Discussion}
\label{sec:discussion}

The preference in the LyS transmission for the Relics simulations over Sherwood supports a picture in which reionization ends as late as $z \sim 5.3$. This is consistent with results from the \lya\ \taueff\ distribution (e.g., \citealt{Zhu2021,Bosman2022, Zhu2022}) and the \mfp\ (e.g., \citealt{Becker2021, Zhu2023, Gaikwad2023, Davies2024}), and indicates that the impact of late-ending reionization is visible in the column density range that dominates the LyS absorption.  

The reionization history impacts the H\,\textsc{i} column density distribution through a combination of factors, including the incidence of neutral islands, the photoionization rate, thermal history, and pressure smoothing. These factors manifest in the LyS transmission, moreover, as an accumulation of opacity at various foreground redshifts.  In the following sections we first deconstruct how the opacity at a given rest-frame wavelength depends on foreground redshift. We then examine the column density distribution of the absorbers and investigate which of these physical processes are primarily responsible for the differences among the models and their varying agreement with the observed quasar composites.

\subsection{Sensitivity to higher-order LyS transitions}
\label{subsec:lys_contribution}

At a fixed quasar rest-frame wavelength, the total LyS absorption receives contributions from multiple transitions arising at different absorber redshifts, with higher-order transitions probing progressively higher redshifts closer to the quasar and remaining sensitive to higher H\,\textsc{i} column densities because of their smaller oscillator strengths. It is important to examine, therefore, the extent to which the difference in total absorption predicted by various models is a reflection of  the differences in higher-order lines near the quasar redshift rather than lower-order lines in the foreground. We adopt $\lambda_{\rm rest}=912$~\AA\ as a reference wavelength and quantify the contribution of the individual LyS transitions. We focus on Sherwood and Relics53, which provide the poorest and best agreement with the observed composites, respectively. The corresponding results at $\lambda_{\rm rest}=890$ and 940~\AA\ are presented in Appendix~\ref{sec:LyStransitions_for_other_wavelengths}.

Fig.~\ref{fig:cumulative_tau_eff_912} shows the cumulative LyS effective optical depth at $\lambda_{\rm rest}=912$~\AA\ for Sherwood and Relics53. Beginning at the highest absorber redshift, the curves build up from right to left as progressively lower-order LyS transitions are included. Quantitatively, transitions from Ly$\delta$ onward contribute approximately $26$--$30\%$ of the total LyS effective optical depth in Sherwood and $31$--$44\%$ in Relics53 across the four composites, demonstrating that higher-order transitions make a substantial contribution to the cumulative LyS absorption.

Perhaps a more important consideration for this work is how the difference in opacity between two models accumulates with redshift. Fig.~\ref{fig:cumulative_delta_tau_eff_912} shows the cumulative difference in effective optical depth between Relics53 and Sherwood. A nonzero difference is already produced by the higher-order LyS transitions and increases as progressively lower-order transitions are added. Transitions from Ly$\delta$ onward account for approximately $60$--$78\%$ of the total $\Delta\tau_{\rm eff}$ across the four composites, highlighting the important contribution of higher-order LyS transitions to the enhanced absorption in Relics53 and its improved agreement with the observed composites.

\subsection{H\,\textsc{i} column density distribution function}

The key quantity governing the LyS transmission is the H\,\textsc{i}
column density distribution function (CDDF) of neutral hydrogen, $f(N_{\rm HI},z)$,
defined as the number of absorbers per unit H\,\textsc{i} column
density and per unit absorption path length, $dX$:
\begin{equation}
f(N_{\rm HI},z)
\equiv
\frac{{\rm d}^2 n}
{{\rm d}N_{\rm HI}\,{\rm d}X}
\equiv
\frac{{\rm d}^2 n}
{{\rm d}N_{\rm HI}\,{\rm d}z}
\frac{H(z)}{H_0}
\frac{1}{(1+z)^2}.
\end{equation}
Following~\cite{Gurvich2017}, we estimate $f(N_{\rm HI},z)$ of each simulation by measuring  the H\,\textsc{i} column density within fixed-width velocity windows of $50\,{\rm km\,s^{-1}}$ along individual simulated lines of sight, and compute the number of absorbers per unit column density and absorption path length (see also \citealt{Rahmati2013,Miller2019}). The choice of velocity-window width and its effect on the inferred CDDF are discussed in Appendix B of \citet{Feron2024}. This approach provides a simple and consistent estimator that can be applied uniformly to all simulations. Because our analysis is based on a finite number of one-dimensional sightlines, however, the estimator becomes increasingly incomplete for high-column-density systems. In particular, the measured $f(N_{\rm HI},z)$ is not expected to accurately recover the absolute abundance of absorbers with $\log(N_{\rm HI}/{\rm cm}^{-2}) \gtrsim 17$, where such systems become increasingly rare and are under-sampled by the available 5000 sightlines (\citealt{Miller2019}). Consequently, we do not attempt to derive quantitative constraints from the $f(N_{\rm HI},z)$ above this threshold. Instead, we use it only for a relative comparison between the simulations, while a robust determination of  the physical distribution at high column densities is beyond the scope of this work. We refer readers to Fig.~5 of \citet{Feron2024} for the H\,\textsc{i} column-density distribution in the Relics57 simulation.

Fig.~\ref{fig:cddf} compares the predicted $f(N_{\rm HI})$ at $z=5.1$, 5.3, 5.6, and 5.9. At $z=5.1$, the differences between the models are relatively modest, but most of Sherwood--Relics simulations still predict a systematically higher abundance of H\,\textsc{i} absorbers than Sherwood at $\log N_{\rm HI}\gtrsim15$. The differences become progressively larger toward higher redshift. At $z=5.6$ and $5.9$, the excess is particularly pronounced at the high-column-density end ($\log N_{\rm HI}\gtrsim17$) in Relics53 and Relics57. Although the absolute $f(N_{\rm HI})$ at these column densities is affected by the limited number of simulated sightlines, the relative excess reflects the presence of residual neutral regions in the radiative-transfer simulations that have not yet been fully ionized. The enhanced abundance of higher-column-density H\,\textsc{i} absorbers in the Sherwood--Relics simulations naturally explains their lower LyS transmission, despite all simulations being tuned to reproduce the same mean \lya\ transmission.

\subsection{Physical origin of the differences in LyS transmission}
\label{subsec:physical_origin}

The Relics53 simulation provides the best agreement with the observed LyS transmission, yielding the lowest $\chi^2$, whereas the Sherwood simulation shows the largest discrepancy. Comparing these two models therefore provides the clearest insight into the physical processes responsible for the differences in the H\,\textsc{i} absorber population and the resulting LyS transmission.

To isolate the origin of the enhanced H\,\textsc{i} absorber population, we perform a series of controlled numerical experiments in which specific physical ingredients are modified while all others are held fixed. These experiments modify the gas densities ($\rho$), peculiar velocities ($v_{\rm pec}$), temperature distribution ($T$), $\Gamma_{\rm HI}$, the assumption of photoionization equilibrium, and the treatment of neutral regions, either individually or in selected combinations. In each experiment, the \lya\ optical depths are rescaled to reproduce the observed \lya\ transmission, allowing us to isolate the effect of the adopted modifications on the LyS absorption. Our main experiments are as follows:

\textbf{Density:} To isolate the effect of the density field, we compare two versions of the simulations in which all other physical ingredients are made to be identical. In both cases, the peculiar velocities are set to zero, the temperature is replaced by a power-law $T$--$\rho$ relation based on temperatures in the Sherwood simulation, the photoionization rate is assumed to be spatially uniform, and the optical depths are recomputed assuming photoionization equilibrium. The only remaining difference between the two models is therefore the underlying gas density field, allowing us to isolate its contribution to the LyS transmission.

\textbf{Temperature:} To isolate the effect of the temperature in ionized regions ($x_{\rm HI}<0.1$), we compare the Sherwood and Relics53 simulations after imposing the same power-law $T$--$\rho$ relation derived from the Sherwood simulation. In this experiment, neutral regions ($x_{\rm HI}>0.1$) are left unchanged and retain their original temperatures.

\textbf{Neutral regions:} To determine the effect of residual neutral regions, we impose photoionization equilibrium using a spatially uniform photoionization rate in the Relics53 simulation, allowing the neutral regions to become ionized. We then additionally replace the temperatures in all regions, including neutral regions, with a power-law $T$--$\rho$ relation. Comparing these results with similar tests restricted to ionized regions allows us to quantify the contribution from residual neutral regions.

Representative figures for these experiments are presented in Appendix~\ref{sec:Numerical_tests}. Here, we summarize the key results from all of the tests performed, using $\lambda_{\rm rest}=912$~\AA\ as the reference wavelength for the quantitative comparisons.

For the lower-redshift composites ($z=5.08$ and $5.32$), the differences between  Sherwood and Relics53 simulations can be almost entirely explained by differences in the density field and the thermal state of the IGM. Approximately 30\% of the difference between the two simulations is attributable to the density field, which reflects the differences in pressure smoothing, while a further $\sim$60\% arises from differences in the temperature distribution. The thermal effect is a consequence of the reionization histories of these simulations. In Sherwood, reionization ended sufficiently long before $z = 6$ that the gas had enough time to thermally relax, producing temperatures that roughly follow a power law in density due to the adiabatic cooling (heating) of underdense (overdense) regions (e.g., \citealt{Hui1997}). This tends to increase (decrease) the recombination rate, and hence increase (decrease) the neutral fraction for a given photoionization rate. In contrast, the later reionization in Relics53 means that the temperature-density relation in ionized regions at $z \lesssim 6$ is generally flatter and exhibits more scatter. Because \lya\ opacity at $5 < z < 6$ tends to be dominated by low-density regions, while LyS opacity regions tend to be dominated by higher densities, the result is that the ratio of LyS/\lya\ opacity in Relics53 tends to be higher than in Sherwood due to the dependence of the recombination rate on temperature. The relationship between H\,\textsc{i} column density and the density and thermal state of the IGM is also illustrated in Fig.~6 of \citet{Feron2024}, who show that absorbers with $N_{\rm HI}\sim10^{14}$--$10^{17}\,{\rm cm}^{-2}$ mostly arise in relatively low-density, ionized gas at $z=5.4$ in the Relics57 simulation.

For the $z=5.64$ composite, about 20\% of the difference between the Sherwood and Relics53 simulations is attributable to differences in the gas density field, while a further $\sim$40\% arises from the thermal structure of the ionized IGM. This thermal contribution reflects the departures of the Relics53 simulation from the nearly power-law $T$--$\rho$ relation followed by the Sherwood simulation. The remaining $\sim$40\% originates from residual neutral regions ($x_{\rm HI}>0.1$)in Relics53.

For the highest-redshift composite ($z=5.93$), differences arising from ionized regions of IGM are smaller. Differences in the gas density field account for only about 10\% of the difference between the Sherwood and Relics53 simulations, while differences in the thermal structure of the ionized IGM contribute a further $\sim$20\%. The remaining $\sim$70\% originates from residual neutral regions.

These results demonstrate that the physical origin of the differences in LyS transmission evolves with redshift. At lower redshifts, where the IGM is almost completely ionized, the differences are primarily determined by the density field and the thermal state of the ionized gas. At higher redshifts, however, residual neutral regions become increasingly important and eventually dominate the difference.

\subsection{Consistency with direct measurements of \mfp}
\label{subsec:mfp_agreement}

\begin{figure}
    \centering
    \includegraphics[width=\linewidth]{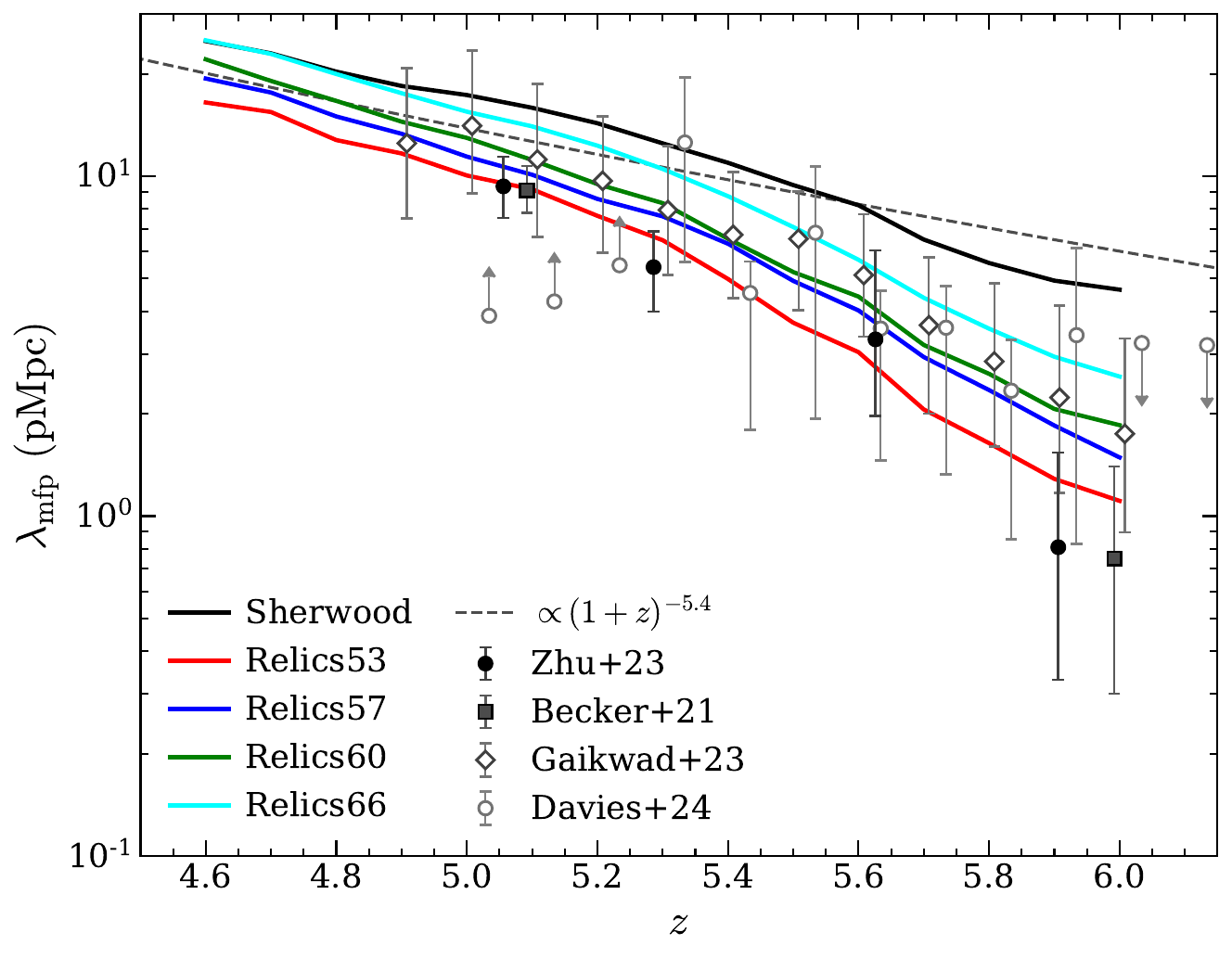}
    \caption{ Evolution of \mfp, predicted by the Sherwood and Sherwood--Relics simulations compared with recent direct measurements from~\citet{Becker2021} and~\citet{Zhu2023} and indirect constraints from~\citet{Gaikwad2023} and~\citet{Davies2024}. The dashed line shows the power-law extrapolation of \mfp\ from lower redshift by \citet{Worseck2014}. Symbols are slightly displaced in redshift for clarity.}
    \label{fig:mfp_sim}
\end{figure}

The higher-order LyS transmission and \mfp\ provide complementary constraints on the \hi\ absorber population. Since the observed higher-order LyS transmission favors the Relics53 simulation, we next test whether the same model is also consistent with previous direct measurements of \mfp\ based solely on LyC transmission. 

Fig.~\ref{fig:mfp_sim} shows the evolution of the predicted \mfp\ for the Sherwood and Sherwood--Relics simulations, measured for each simulation following the procedure described in Section~\ref{subsec:methods}, together with recent direct observational measurements (\citealt{Becker2021, Zhu2023}) and indirect constraints from the \lya\ \taueff\ distribution (\citealt{Gaikwad2023, Davies2024}). The Relics53 simulation, which provides the best agreement with the observed higher-order LyS transmission, also provides the best match to the evolution of the observed direct measurements of \mfp\ over $5\lesssim z\lesssim6$. In contrast, models with substantially earlier reionization generally predict a longer \mfp\ at these redshifts, reflecting the reduced abundance of high-column-density \hi\ absorbers after the IGM has had more time to ionize and dynamically relax. This agreement is consistent with the findings of \citet{Feron2024}, who showed that Sherwood--Relics simulations with late-ending reionization can reproduce the rapid evolution of the observed \mfp\ at $5\lesssim z\lesssim6$.

\subsection{Implications for the end of reionization}

The preferred Relics53 model corresponds to a relatively late reionization history, in which the volume-averaged neutral fraction approaches $10^{-3}$ at $z=5.3$ \citep{Puchwein2023}.
Although our simulations sample only a discrete set of reionization histories, this result is consistent with a growing body of observational evidence suggesting that reionization may have extended until $z\sim5.3$. In particular, recent studies based on \lya\ forest opacity fluctuations~\citep{Bosman2018, Eilers2018,Kulkarni2019,Bosman2022}, quasar damping-wing measurements~\citep{Becker2024, Zhu2024}, dark-gap and dark pixel statistics~\citep{Zhu2021, Zhu2022, Davies2026}, the evolution of  metal absorbers near $z\sim6$ (e.g.\ \citealt{Becker2019,Cooper2019, DOdorico2022}), the association of large-scale galaxy underdensities with extended dark \lya\ troughs (\citealt{Becker2018, Kashino2020, Christenson2023, Conaboy2026,Gangolli2025, Zhu2026}), and the evolving  \mfp ~\citep{Becker2021,Cain2021,Zhu2023,Gaikwad2023,Davies2024,Roth2024, Satyavolu2024} have all pointed toward a relatively late end to reionization. Our results provide additional support for this scenario. 
Independent measurements of broadband LyS transmission
toward 279 galaxies at $5.45<z<6.85$ using
JWST/NIRCam F070W imaging also show a preference for
the late-reionization Sherwood--Relics model over a
homogeneous-UVB model (Zhu et al., in preparation). These findings are consistent
with our results, despite the different observational
approaches and the possible contribution of absorption
associated with the source galaxies.

The preference in our analysis for a late reionization history
is robust to uncertainties in the ionization-efficiency
parameter $\xi$. Although varying $\xi$ changes the predicted
transmission and the absolute $\chi^2$ values, the
Sherwood--Relics models remain strongly preferred over
Sherwood across the full range considered. For our fiducial
value, $\xi=0.67$, Relics53 provides the lowest combined
reduced $\chi^2$, $\chi^2_\nu=1.23$, compared with $1.96$
for Sherwood. Relics53 is also preferred for $\xi=0.33$
and $\xi=1.0$, with $\chi^2_\nu=1.23$ and $1.25$,
respectively. Overall, the preference for Relics53 and
a relatively late reionization history remains insensitive
to the adopted value of $\xi$ over the range considered here.

More broadly, our results demonstrate that higher-order LyS absorption provides a powerful and complementary probe of the late stages of reionization. Although the simulations are rescaled to reproduce the observed \lya\ transmission, they predict measurably different higher-order LyS absorption because these transitions retain sensitivity to a broader range of \hi\ column densities. Consequently, higher-order LyS absorption provides additional discriminatory power for distinguishing between different reionization histories beyond that available from \lya\ transmission alone.

\section{Summary and conclusion }
\label{sec:summary}

Higher-order LyS absorption provides a  bridge between the low-column-density \hi\ traced by the \lya\ forest and the higher-column-density absorbers that strongly influence the ionizing photon mean free path, offering a broader view of the ionization structure of the IGM near the end of reionization.
In this work, we used four composite spectra centered at redshifts $z=5.08$, 5.32, 5.64, and 5.93 constructed from a sample of 100 quasar spectra spanning the redshift range $4.9 \leq z \leq 6.1$ obtained with the Keck/LRIS, Keck/ESI, Gemini/GMOS, and VLT/X-Shooter spectrographs. We compared these observations to realistic forward-modeled composite spectra generated from simulations with a range of reionization histories. These included the Sherwood simulation, which assumes an early, homogeneous reionization, and four Sherwood--Relics radiative-transfer simulations that incorporate patchy reionization with different hydrogen reionization completion redshifts. By comparing the observed and simulated higher-order LyS transmission, we investigated both the physical origin of the absorption and its implications for the timing of the final stages of cosmic reionization. Our main findings are summarized as follows.

\begin{enumerate}

\item The Sherwood simulation, which assumes an early, homogeneous reionization and a spatially uniform UVB, is unable to reproduce the observed higher-order LyS transmission in any of the four quasar composites. In contrast, the Sherwood--Relics radiative-transfer simulations, which incorporate patchy reionization and spatial fluctuations in the ionization state of the IGM, provide substantially better agreement with the observations. 

\item Among the Sherwood--Relics models considered in this work, the Relics53 simulation, which has hydrogen reionization ending at $z=5.3$, provides the best match to the observed quasar composites. This result is in good agreement with an increasing number of observational studies that favor a late completion of reionization.

\item  The physical origin of the enhanced higher-order LyS absorption in the late-ending Relics runs compared to Sherwood evolves with redshift. At $z=5.08$ and $5.32$, the larger abundance of H\,\textsc{i} absorbers in the preferred Relics53 model is fully explained by differences in the temperature--density relation and the gas density field, reflecting reduced pressure smoothing and elevated temperatures following a later reionization history. At $z=5.64$ and $5.93$, however, these effects alone cannot account for the observed absorption, and residual neutral islands make an essential contribution.

\item The evolution of \mfp\ in the Relics53 simulation is consistent with previous direct measurements of \mfp\ over $5 < z < 6$ based on LyC transmission alone~\citep{Becker2021, Zhu2023}.

Finally, our results establish higher-order LyS absorption as a powerful probe of the \hi\ absorber population. Larger quasar samples will enable better constraints on the abundance and evolution of neutral hydrogen absorbers during and after the final stages of reionization.

\end{enumerate}

\section*{Acknowledgements}

We thank Anson D'Aloisio for helpful discussions. S.H., G.D.B, and H.H. were supported by JWST Program 4092. Y.Z. acknowledges support from the NIRCam Science Team contract to the University of Arizona, NAS5-02105. Y.Z. is also supported by JWST Program \#6434. JSB acknowledges support from STFC consolidated grant ST/X000982/1. Support for programs 4092 and 6434 was provided by NASA through a grant from the Space Telescope Science Institute, which is operated by the Association of Universities for Research in Astronomy, Inc., under NASA contract NAS 5-03127.  

Some of the data presented herein were obtained at the  W.M. Keck Observatory, which is operated as a scientific partnership between the California Institute of Technology and the University of California; it was made possible by the generous support of the W.M. Keck Foundation. The authors wish to recognize and acknowledge the very significant cultural role and reverence that the summit of Maunakea has always had within the indigenous Hawaiian community. We are most fortunate to have the opportunity to conduct observations from this mountain.  This research also made use of the Keck Observatory Archive (KOA), which is operated by the W.M. Keck Observatory and the NASA Exoplanet Science Institute (NExScI), under contract with the National Aeronautics and Space Administration.  Further observations were made with ESO telescopes at the La Silla Paranal Observatory under program IDs 60.A-9024, 084.A-0390, 085.A-0299, 086.A-0162, 086.A-0574, 088.A-0897, 091.C-0934,
097.B-1070, and 098.B-0537, and 1103.A-0817.  This work is also based on observations obtained at the international Gemini Observatory, a program of NSF’s NOIRLab, which is managed by the Association of Universities for Research in Astronomy (AURA) under a cooperative agreement with the National Science Foundation. on behalf of the Gemini Observatory partnership: the National Science Foundation (United States), National Research Council (Canada), Agencia Nacional de Investigaci\'{o}n y Desarrollo (Chile), Ministerio de Ciencia, Tecnolog\'{i}a e Innovaci\'{o}n (Argentina), Minist\'{e}rio da Ci\^{e}ncia, Tecnologia, Inova\c{c}\~{o}es e Comunica\c{c}\~{o}es (Brazil), and Korea Astronomy and Space Science Institute (Republic of Korea).

The hydrodynamical simulations used in this work were performed using the Joliot Curie supercomputer at the Tr{\'e}s Grand Centre de Calcul (TGCC) and the Cambridge Service for Data Driven Discovery (CSD3), part of which is operated by the University of Cambridge Research Computing on behalf of the STFC DiRAC HPC Facility (\href{https://www.dirac.ac.uk/}{www.dirac.ac.uk}). We acknowledge the Partnership for Advanced Computing in Europe (PRACE) for awarding us time on Joliot Curie in the 16th call. The DiRAC component of CSD3 was funded by BEIS capital funding via STFC capital grants ST/P002307/1 and ST/R002452/1 and STFC operations grant ST/R00689X/1. This work also used the DiRAC@Durham facility managed by the Institute for Computational Cosmology on behalf of the STFC DiRAC HPC Facility. The equipment was funded by BEIS capital funding via STFC capital grants ST/P002293/1 and ST/R002371/1, Durham University and STFC operations grant ST/R000832/1. DiRAC is part of the National e-Infrastructure.

This work made use of the following software packages: numpy~\citep{Harris2020}, matplotlib~\citep{Hunter2007},
scipy~\citep{Virtanen2020}, astropy, a community-developed core
Python package for Astronomy~\citep{Astropy2013,Astropy2018}.




\bibliographystyle{mnras}
\bibliography{lys_ref} 



\appendix

\section{Effect of the Quick \lya\ Approximation on LyS Transmission}
\label{sec:quick_lya_effects}

\begin{figure*}
    \centering
    \includegraphics[width=\linewidth]{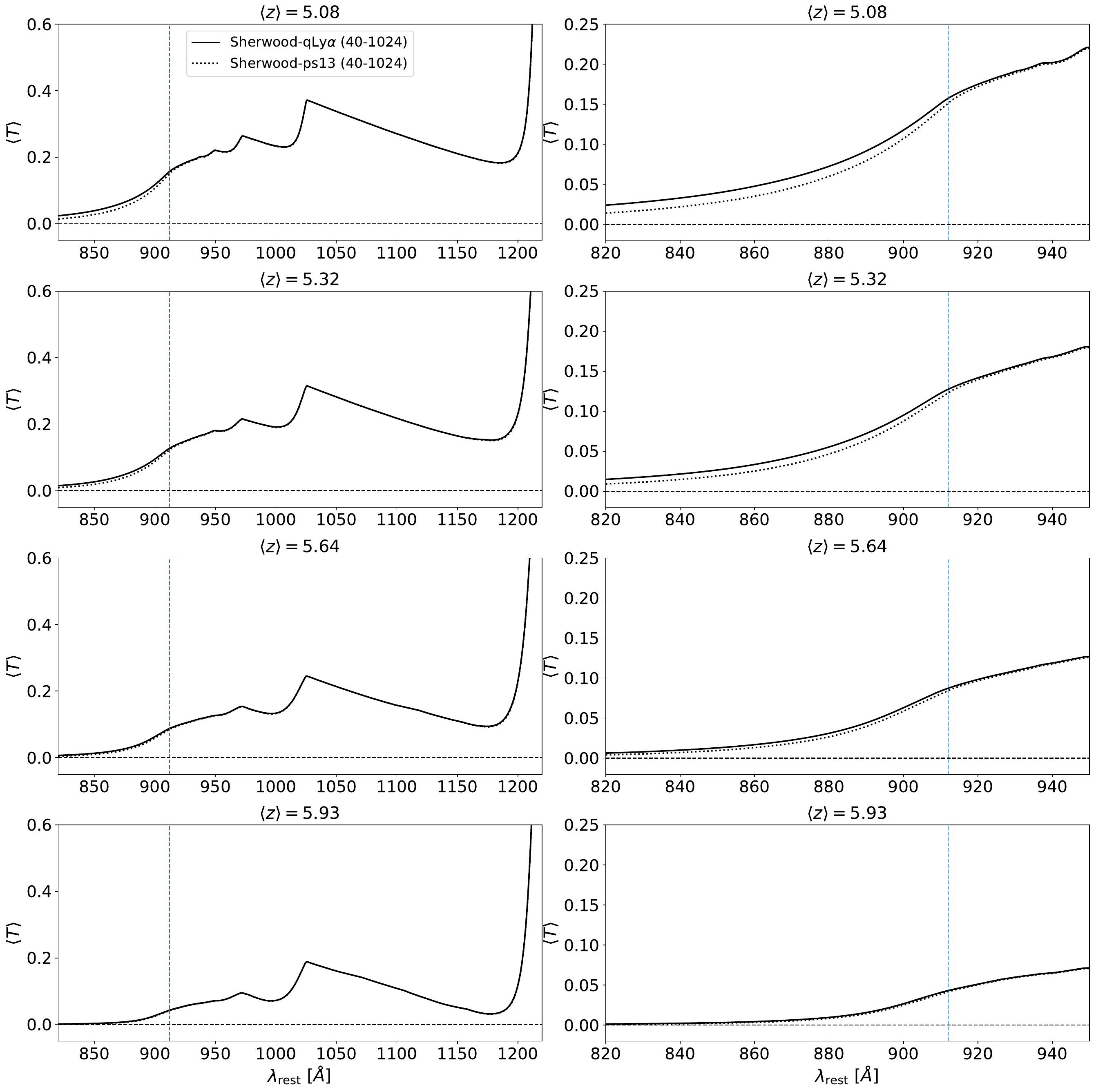}
\caption{Comparison of the mean LyS transmission predicted by two simulations with different treatments of dense regions. Sherwood-q\lya\ (solid) employs the quick-\lya\ approximation used in this work, in which gas with overdensity $\Delta>1000$ and temperature $T<10^5$ K is converted into collisionless star particles. Sherwood-ps13 (dotted) instead uses the star formation and feedback model of ~\citet{Puchwein2013}, retaining high-density gas in halos. A self-shielding correction following~\citet{Chardin2018} is also applied to the Sherwood-ps13 model. The left column shows the full wavelength range, while the right column presents a zoom-in around the Lyman-limit region.
}
    \label{fig:quick_lys_test}
\end{figure*}

\begin{figure*}
    \centering
    \includegraphics[width=\linewidth]{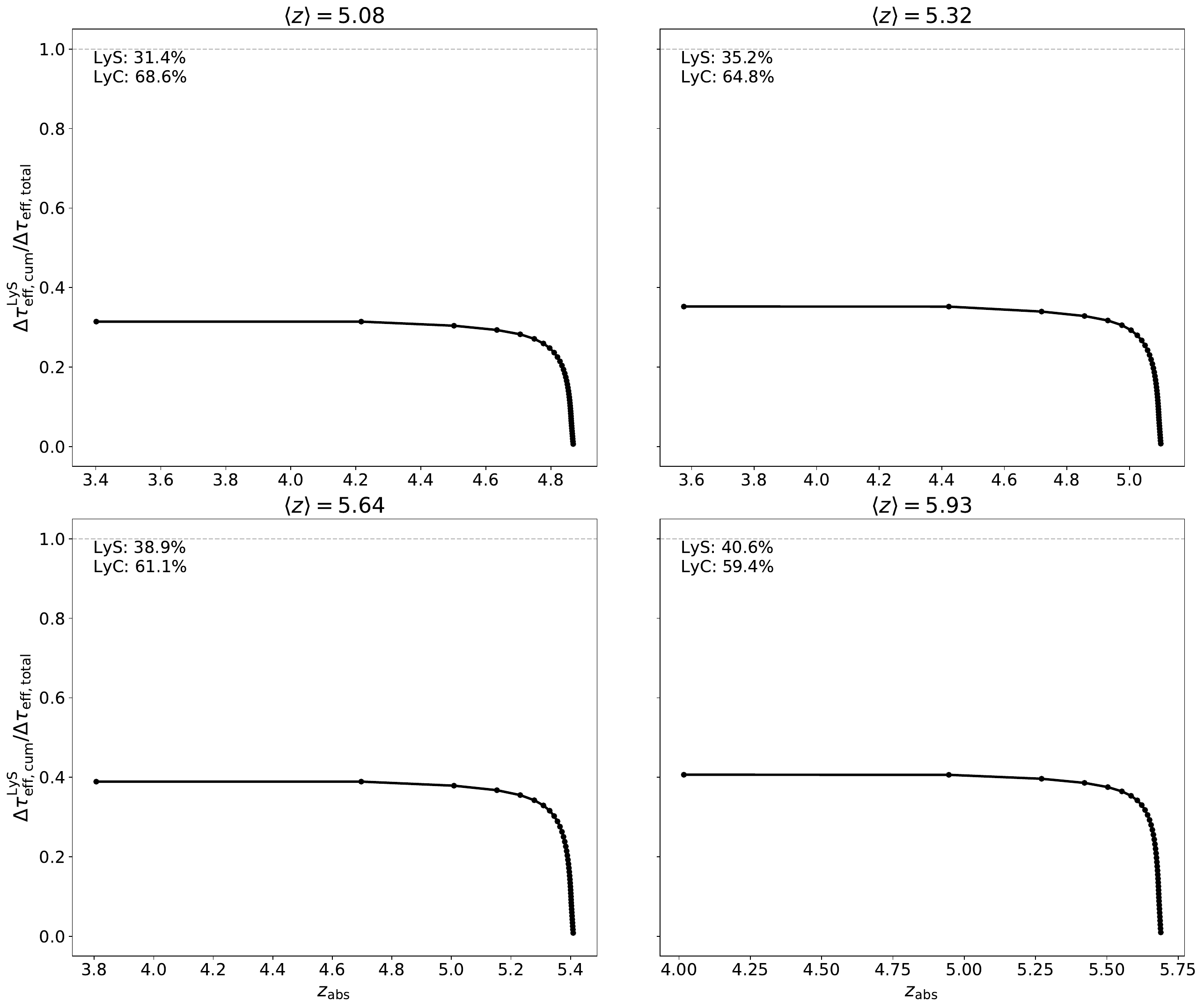}
\caption{
Cumulative contribution of LyS absorption to the difference
in total effective optical depth between Sherwood-q\lya\
and Sherwood-ps13 at 880~\AA, as a function of absorber
redshift. The cumulative difference is constructed by
successively adding the contributions of individual LyS
transitions, beginning at the highest absorber redshift,
closest to the quasar, and proceeding toward lower redshift.
The points mark the contributions of individual LyS
transitions along the cumulative curve. The annotated
percentages indicate the relative contributions of LyS
and LyC absorption to the total difference in effective
optical depth.
}
\label{fig:quick_lys_fraction}
\end{figure*}

Here, we investigate the effect of the quick-\lya\ approximation used in the Sherwood and Sherwood–Relics simulations on the predicted LyS transmission.  
For this test, we use the 40–1024 and 40–1024-ps13 models listed in Table~1 of \citet{Bolton2017}, which we refer to as Sherwood-q\lya\ and Sherwood-ps13, respectively. Both have the same box size as our fiducial 40–2048 model, but a dark matter particle mass eight times larger. The key difference between these two simulations is their treatment of dense gas: Sherwood-q\lya\ employs the same quick-\lya\ approximation adopted in this work, whereas Sherwood-ps13 uses the star formation and energy-driven outflow model of ~\cite{Puchwein2013}. For the Sherwood-ps13 model, the self-shielding of dense gas is additionally accounted for by correcting the neutral hydrogen fractions following~\citet{Chardin2018}. Since the two simulations otherwise share the same numerical setup, comparing their LyS transmission predictions provides an indication of the sensitivity to the dense, cool gas omitted by the quick-\lya\ approximation.

The results of the comparison are displayed in
Fig.~\ref{fig:quick_lys_test}. The mean transmission predicted by
Sherwood-ps13 is slightly lower than that of Sherwood-q\lya\
across all four redshift bins. At 912~\AA, the differences are
only about 3--4 per cent, while at 880~\AA\ they increase to
approximately 13--19 per cent. The differences at 880~\AA\
decrease with increasing redshift, from about 18 per cent at
$z=5.08$ to 13 per cent at $z=5.93$. To further investigate the origin of these differences, we
decompose the change in total effective optical depth at
880~\AA\ into contributions from LyS and LyC absorption,
as shown in Fig.~\ref{fig:quick_lys_fraction}.
We find that LyS absorption accounts for approximately
35--42 per cent of the difference across the four redshift
bins, while LyC absorption contributes the remaining
58--65 per cent.

Assuming that similar results apply to all of our models, the quick-\lya\ approximation would tend to have a moderate impact on our results over the LyC region, but mainly for the two lower-redshift composites.  Our LyS results should be largely unaffected.

\section{Effect of Correlations on the Mean Transmission}
\label{app:correlations}

\begin{figure*}
    \centering
    \includegraphics[width=\linewidth]{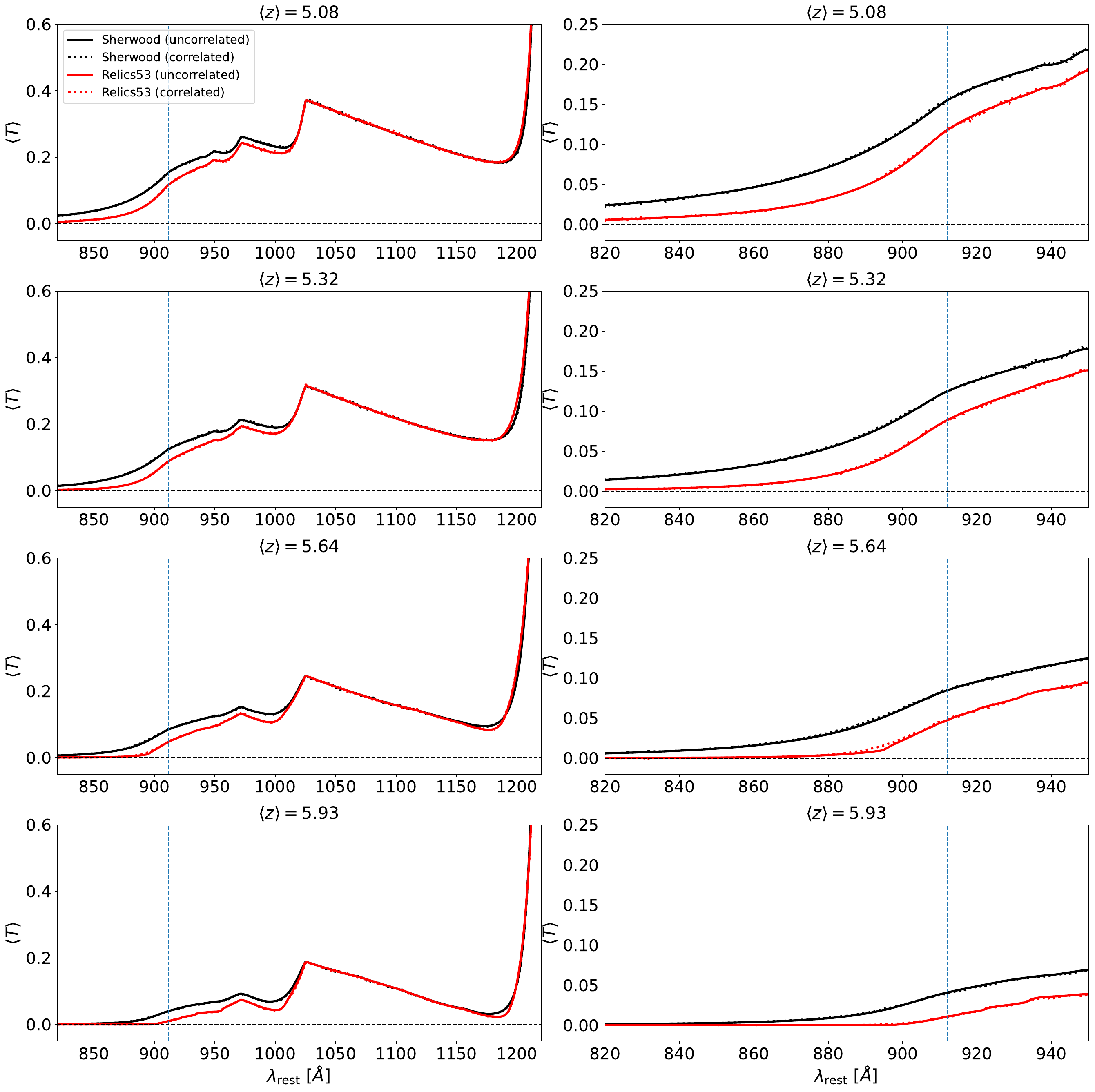}

\caption{
Comparison of the uncorrelated (solid) and correlated (dotted)
mean transmission predictions for Sherwood (black) and
Relics53 (red), using approximately $10^4$ simulated
sightlines at each redshift. The left panels show the full
wavelength range, while the right panels provide a closer
view of the Lyman-limit region.
}
\label{fig:correlation_test}
\end{figure*}

In our fiducial model, the mean transmission is obtained by multiplying the mean transmissions of the individual LyS transitions and LyC absorption, thereby neglecting possible correlations between these components:
\begin{equation}
\langle T(\lambda) \rangle_{\rm uncorr}
=
\langle T_{\rm LyC}(\lambda) \rangle
\prod_j \langle T_j(\lambda) \rangle.
\end{equation}
Here, $T_j$ represents the transmission from the $j$th LyS
transition, and $T_{\rm LyC}$ denotes the LyC transmission.
The product extends over all LyS transitions included in
our calculation.

To assess the accuracy of this approximation, we alternatively calculate the total transmission directly by multiplying the transmissions of the LyS transitions and LyC absorption along individual sightlines.  The mean total transmission is then computed as
\begin{equation}
\langle T(\lambda) \rangle_{\rm corr}
=
\left\langle
T_{\rm LyC}(\lambda)
\prod_j T_j(\lambda)
\right\rangle_{\rm LOS},
\end{equation}
where the average is taken over simulated sightlines.
This approach naturally preserves correlations between the
different absorption components.

For this test, we generate 10000 simulated lines of sight for each quasar composite, adopting the mean redshift and $R_{\rm eq}$ of the corresponding composite. We construct extended lines of sight by stitching together simulated skewers across the relevant redshift range, applying the same optical-depth rescaling and quasar proximity-zone treatment as in our fiducial model (see section~\ref{subsec:methods}). The LyS optical depths are obtained directly from the simulated spectra, while the LyC optical depth is calculated from the neutral hydrogen column densities along each sightline. The resulting transmission spectra are convolved to a common instrumental resolution, rebinned, and averaged over all sightlines to obtain the mean transmission. We have verified that the mean transmission is well converged with 10000 sightlines and that the agreement between the two approaches is insensitive to the adopted spectral resolution and wavelength binning.

Figure~\ref{fig:correlation_test} compares the mean transmission obtained from the two approaches for Sherwood and Relics53 at all four composite redshifts. We find that the resulting transmission profiles agree closely across the entire wavelength range considered, including wavelengths below the Lyman limit where both LyS and LyC absorption contribute. This agreement suggests that correlations between the different absorption components have only a small effect on the predicted mean transmission, supporting the approximation adopted in our fiducial model.

This result can be understood in terms of the strengths of the lines that contribute to the absorption at a given wavelength. Absorption from stronger, lower-order LyS lines will arise from foreground redshifts that are widely separated, and thus will be intrinsically uncorrelated.  Absorption from lines with rest wavelengths nearer to 912~\AA\ will trace more similar redshifts and will therefore be increasingly correlated; however, the absorption from these lines in our models is weak enough that the correlations do not significantly impact the mean transmission.  We note that this result may only apply to optically thin models \citep[see also][]{Becker2021,Zhu2023}.  In models that contain optically thick absorbers, correlations among LyS lines and between LyS and LyC must be taken into account in order to estimate the mean transmission (e.g.,\ \citealt{Roth2024}; Gaikwad et al., in prep).  The Relics53 simulation used here does contain optically thick neutral islands; however,  our treatment of the proximity zones largely removes them from the vicinity of the QSOs (Section \ref{Treatment-of-Neutral-Regions}).

\section{Alternative Transmission Calculations Approach} 

\label{sec:Alternative_Transmission_Calculations}

\begin{figure*}
    \centering
    \includegraphics[width=\linewidth]{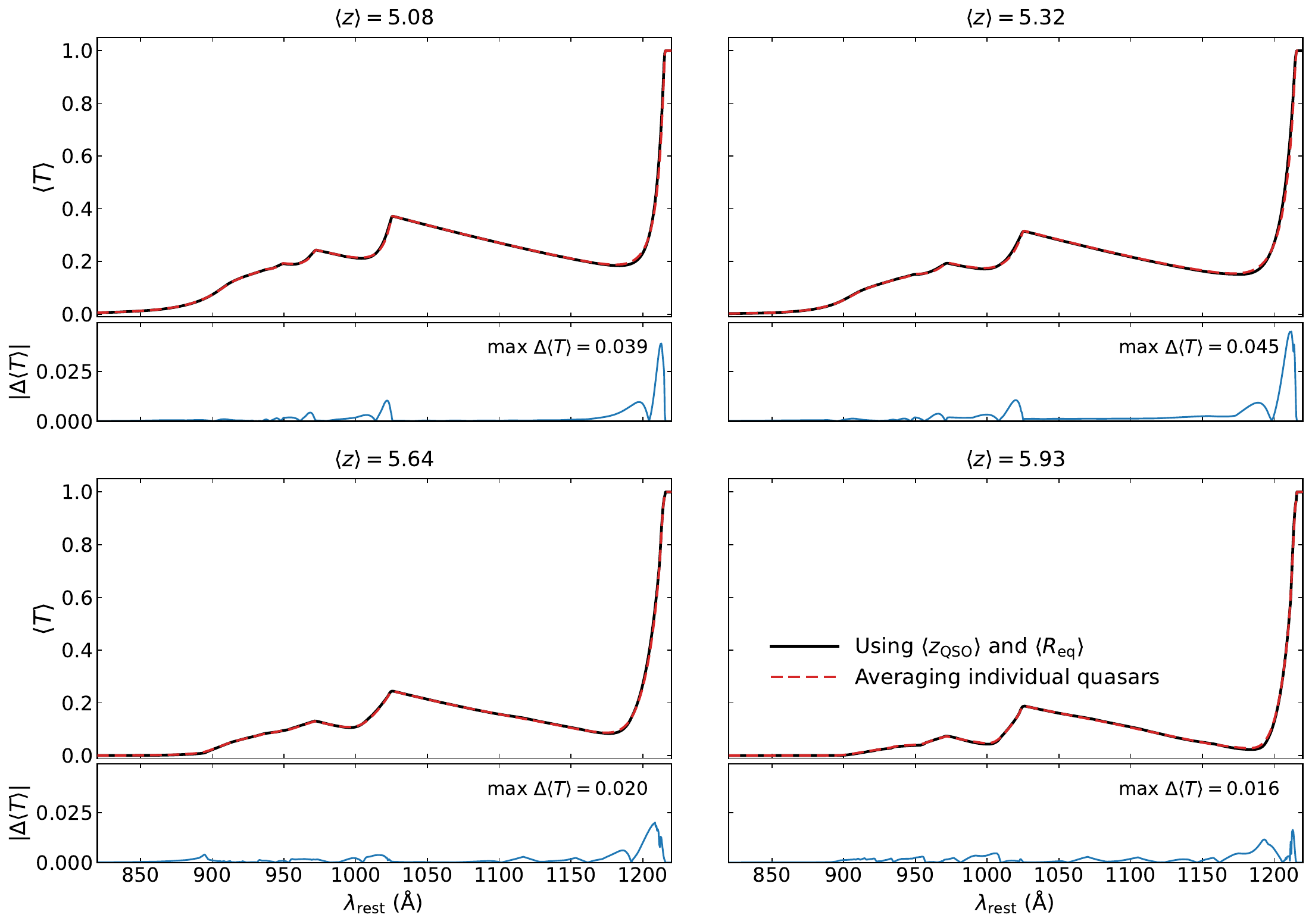}
\caption{
Comparison of the mean transmission profiles obtained using two different approaches for the four quasar composites for the Relics53 simulation. In the first approach (black solid curves), the transmission is computed using the mean composite redshift, $\langle z_{\rm QSO}\rangle$, and the simulation-dependent mean $R_{\rm eq}$, $\langle R_{\rm eq}\rangle$, for each composite. In the second approach (red dashed curves), the transmission profile is calculated separately for every quasar using its individual redshift and simulation-dependent $R_{\rm eq}$, and the resulting transmission profiles are then averaged within each composite. The lower panels show the absolute difference, $|\Delta\langle T\rangle|$, between the two methods.
}
    \label{fig:req_test}
\end{figure*}

\begin{figure*}
    \centering
    \includegraphics[width=\linewidth]{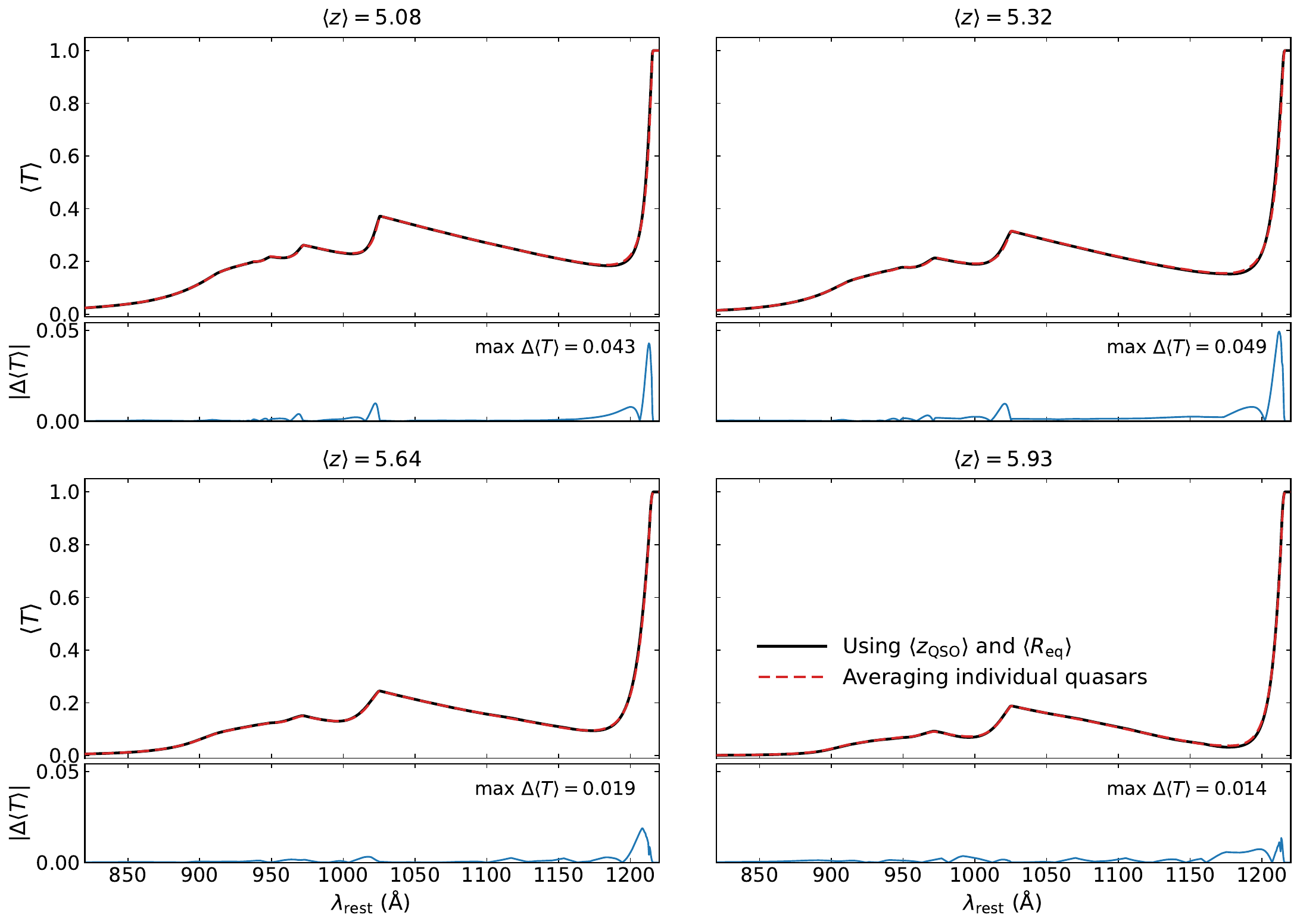}
\caption{Like Fig.~\ref{fig:req_test}, but for Sherwood simulation.
}
    \label{fig:req_test2}
\end{figure*}

Throughout this work, the transmission profile for each quasar composite is computed using the mean quasar redshift and the mean simulation-dependent $R_{\rm eq}$. This approximation requires only a single transmission calculation for each simulation at each composite redshift, making it substantially more computationally efficient than computing the transmission profile separately for every quasar. To assess the accuracy of this approximation, we compute the transmission profile for each individual quasar using its own redshift and simulation-dependent $R_{\rm eq}$, and then average the resulting transmission profiles within each composite. Figs.~\ref{fig:req_test} and~\ref{fig:req_test2} compare the two approaches for the Relics53 and Sherwood simulations, respectively.

For both the Relics53 and Sherwood simulations, the two approaches produce nearly identical transmission profiles over the full wavelength range considered and for all four quasar composites. In particular, no appreciable differences are found within the $820$--$950$~\AA\ interval used for the model comparison. This demonstrates that using the mean quasar redshift and mean simulation-dependent $R_{\rm eq}$ for each composite provides an accurate approximation to averaging the transmission profiles computed for the individual quasars. We therefore adopt the composite-average approach throughout this work.

\section{Sensitivity to the extent of quasar-ionized regions}
\label{sec:Sensitivity_to_the_Definition_of_the Quasar-affected Region}

In the main analysis, we assume that quasars will locally ionize the IGM out to a distance at which the quasar photoionization rate (Equation~\ref{eq:Gamma_qso_r}) equals the  background photoionization rate, i.e.,
$(\Gamma_{\rm QSO}/\Gamma_{\rm bg})_{\rm th}=1$.
We investigate how the predicted transmission depends on this choice by considering threshold values of
$(\Gamma_{\rm QSO}/\Gamma_{\rm bg})_{\rm th}=0.5$, 1, 2, and 4.

\begin{figure*}
    \centering
    \includegraphics[width=\linewidth]{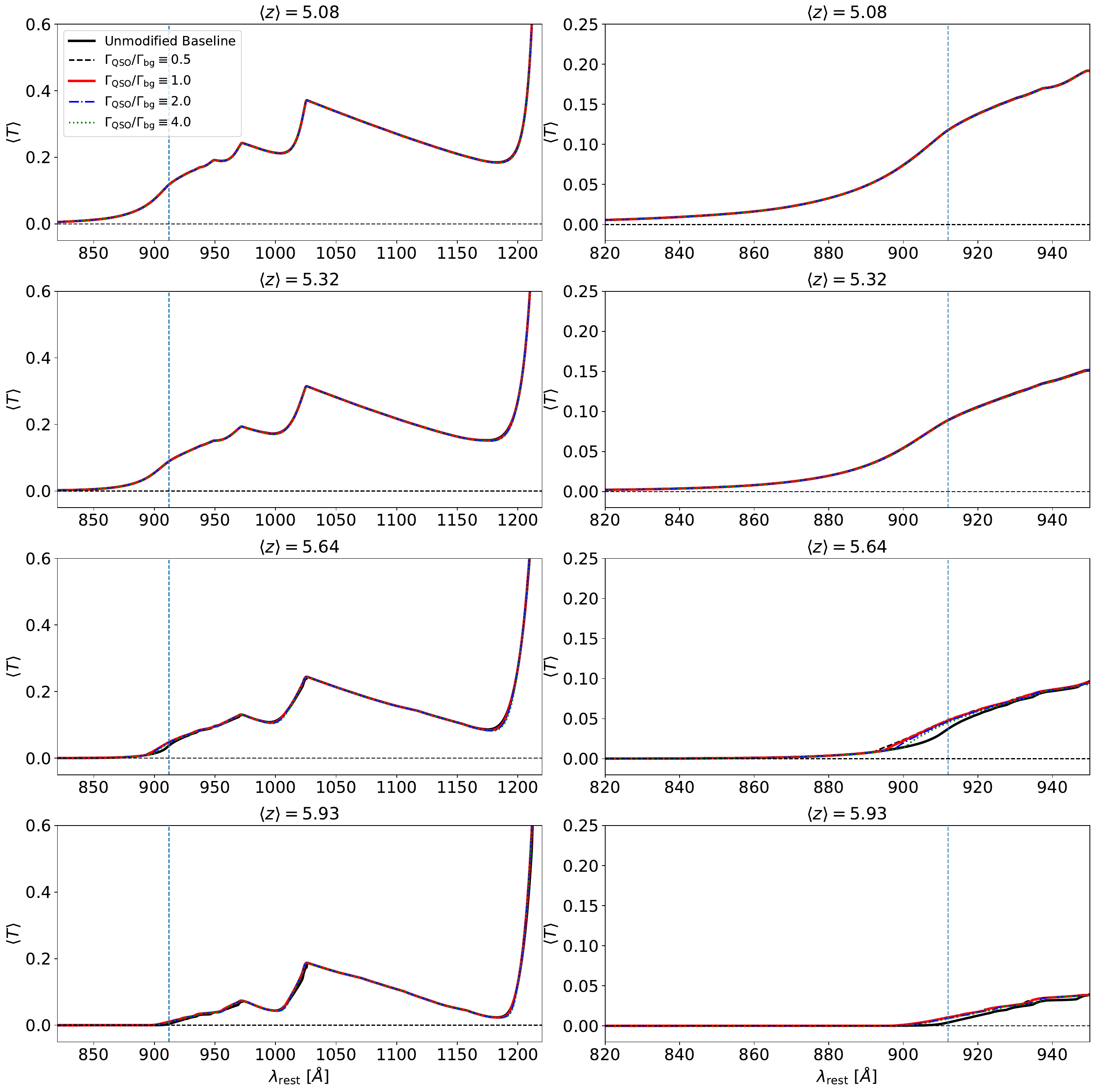}
\caption{
Comparison of the predicted mean transmission profiles obtained using different definitions of the extent of the quasar-affected region in the Relics53 simulation. The black curve shows the original Relics53 simulation, while the colored curves show models in which the uniform UVB equilibrium simulation is adopted out to the distance where $\Gamma_{\rm QSO}/\Gamma_{\rm bg}=0.5$, 1, 2, and 4. The left column shows the full wavelength range, while the right column presents a zoom-in around the Lyman-limit region.
}
    \label{fig:ionization_region_test}
\end{figure*}

Fig.~\ref{fig:ionization_region_test} shows the resulting transmission profiles for the Relics53 simulation, which is expected to exhibit the greatest sensitivity to this choice because it has the latest reionization history among the models considered. As expected, the transmission profiles of the two lower-redshift composites, whose mean redshifts lie below the reionization completion redshift of the simulation, are essentially unchanged for all values of $\Gamma_{\rm QSO}/\Gamma_{\rm bg}$, since the IGM is already highly ionized and contains no significant residual neutral regions. For the two higher-redshift composites, the predicted transmission varies only weakly with the adopted transition criterion. The largest differences occur near the Lyman limit, where the precise extent of the quasar-affected region has the greatest influence. At shorter wavelengths, however, the transmission profiles are nearly identical for all cases. We therefore conclude that our results are insensitive to the exact choice of the transition distance, and adopting the fiducial value $\Gamma_{\rm QSO}/\Gamma_{\rm bg}=1$ does not affect the conclusions of this work.

\section{Transmission Profiles and Effective Optical Depth for Alternative Values of $\xi$}
\label{sec:Transmission_Profiles_for_Alternative_Values}

\begin{figure*}
    \centering
    \includegraphics[width=\linewidth]{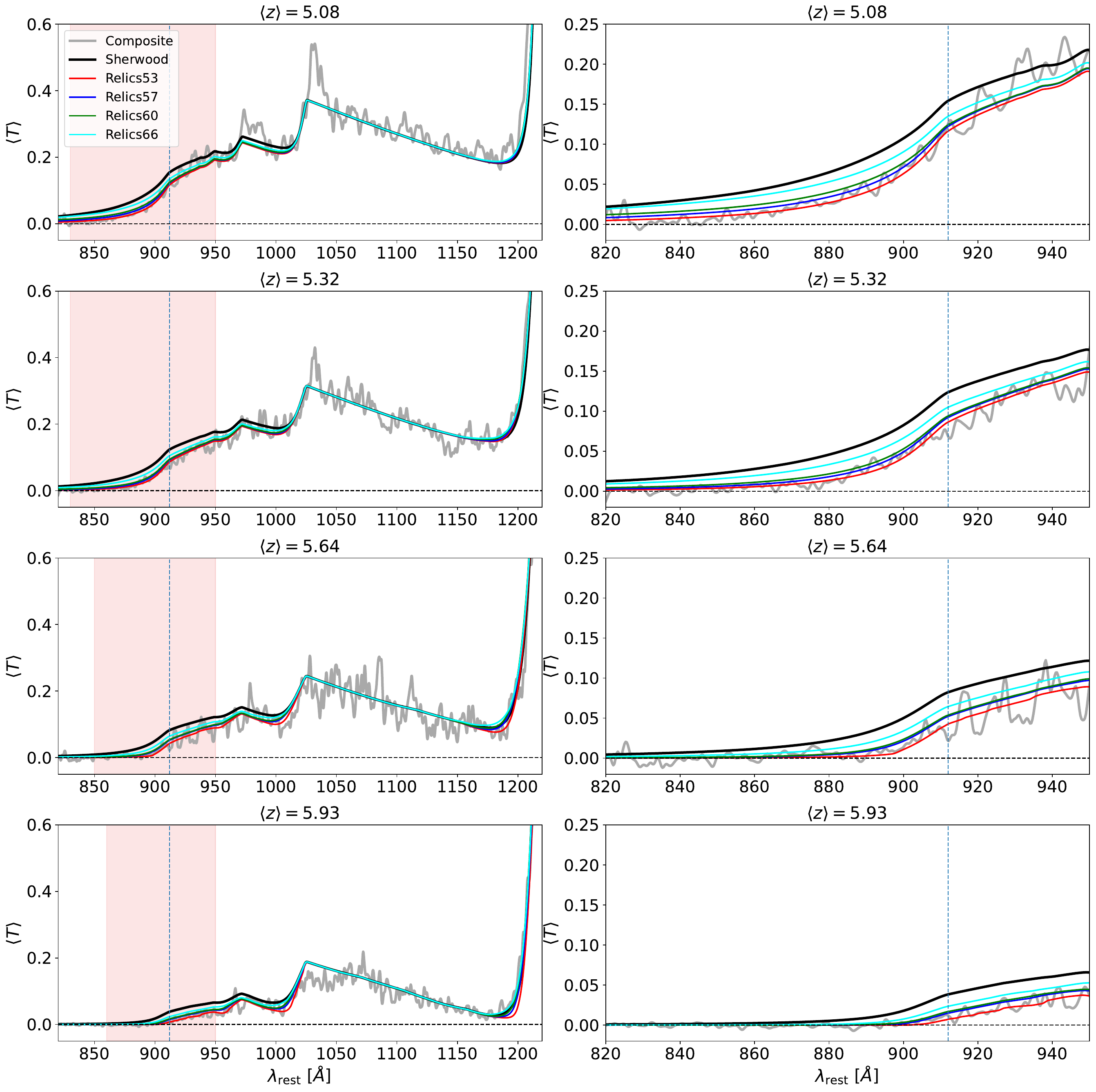}
\caption{Same as Fig.~\ref{fig:transmissions_zeta67}, but assuming $\xi=0.33$.}
\label{fig:transmissions_zeta33}
\end{figure*}

\begin{figure*}
    \centering
    \includegraphics[width=\linewidth]{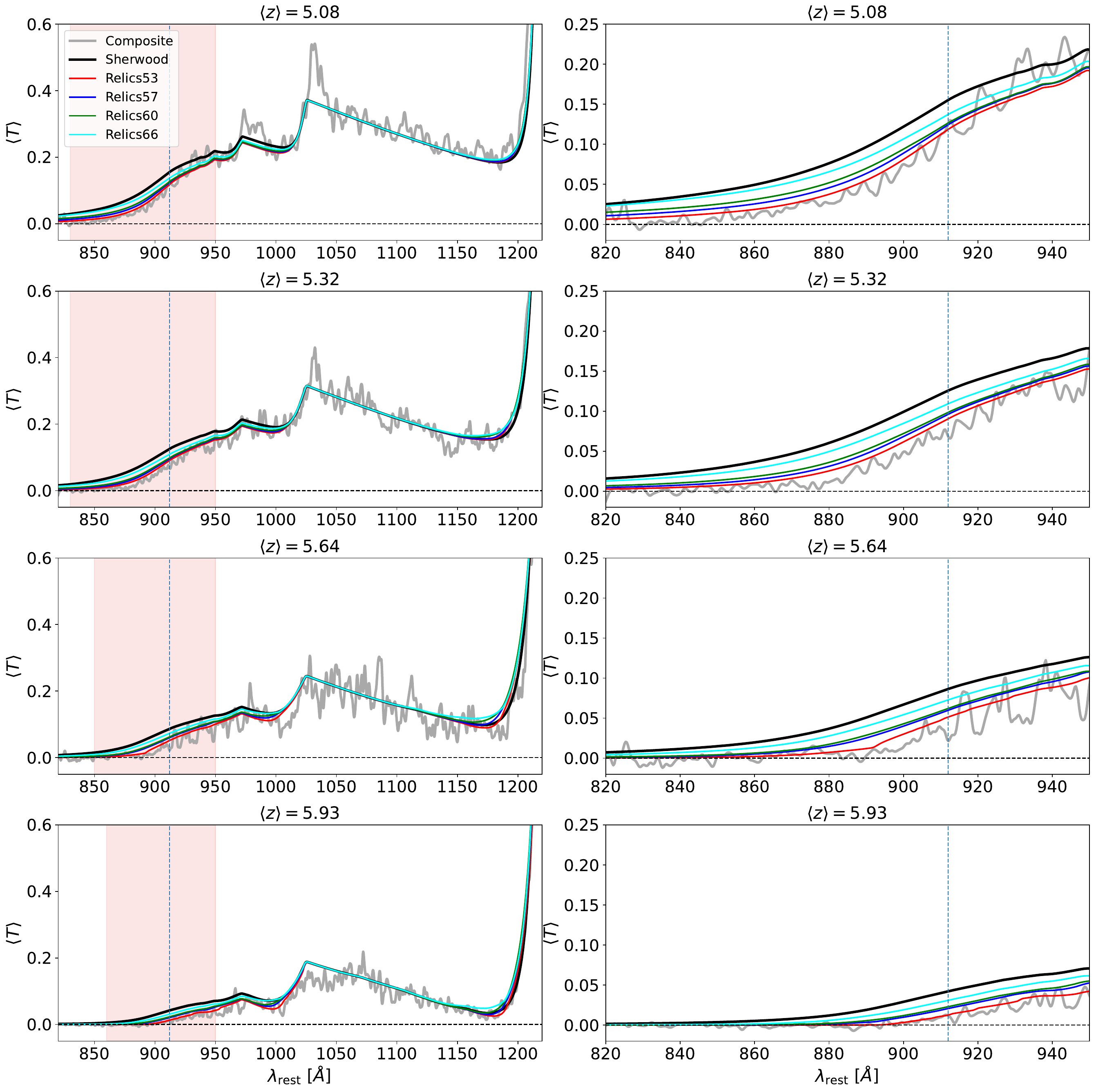}
\caption{Same as Fig.~\ref{fig:transmissions_zeta67}, but for $\xi=1.0$.}
\label{fig:transmissions_zeta100}
\end{figure*}

Figs.~\ref{fig:transmissions_zeta33} and
\ref{fig:transmissions_zeta100} show the predicted transmission profiles
obtained using $\xi=0.33$ and $\xi=1.0$, respectively. These figures are
presented in the same format as the fiducial $\xi=0.67$ results shown in
Fig.~\ref{fig:transmissions_zeta67}, allowing the effect of varying
$\xi$ to be assessed directly across the four redshift composites and
 different simulations. Changing $\xi$ modifies the extent to which
the enhanced quasar ionizing field reduces the surrounding absorption,
and therefore alters the predicted transmission. The corresponding
 $\chi^2_\nu$ values are listed in
Table~\ref{tab:chi2_xi_comparison}.

\begin{figure*}
    \centering
    \includegraphics[width=0.9\textwidth]
    {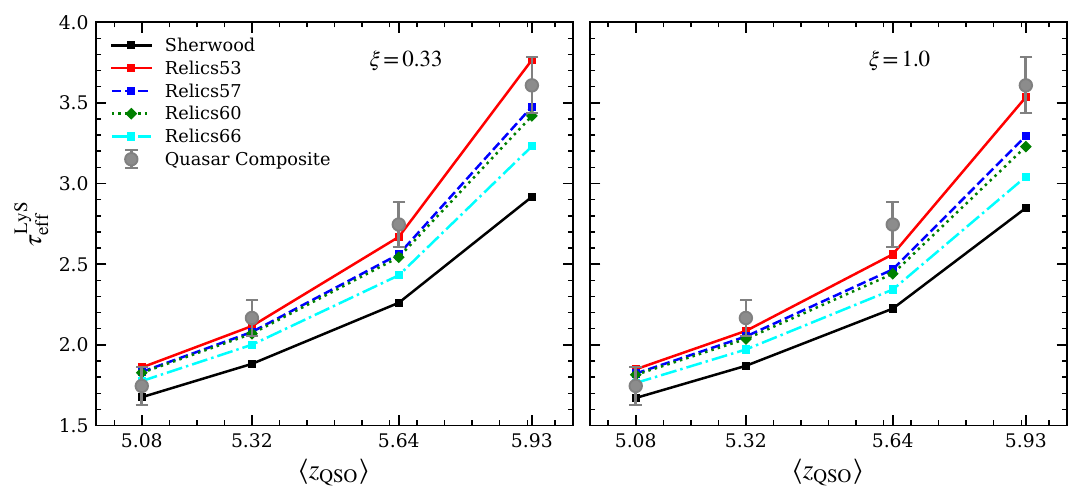}
    \caption{
    Same as Fig.~\ref{fig:tau_eff_lys}, but for
    $\xi=0.33$ (left) and $\xi=1.0$ (right).
        }
    \label{fig:tau_eff_lys_appendix}
\end{figure*}

Figure~\ref{fig:tau_eff_lys_appendix} shows the corresponding
evolution of $\tau_{\rm eff}^{\rm LyS}$ over 912--950~\AA\
for $\xi=0.33$ and $\xi=1.0$, in the same format as the
fiducial results shown in Fig.~\ref{fig:tau_eff_lys}.
The Sherwood--Relics models provide better overall agreement
with the observed LyS effective optical depths than Sherwood
for both values of $\xi$.

\section{Contributions of individual LyS transitions at other wavelengths}
\label{sec:LyStransitions_for_other_wavelengths}

\begin{figure*}
    \centering
    \includegraphics[width=\linewidth]{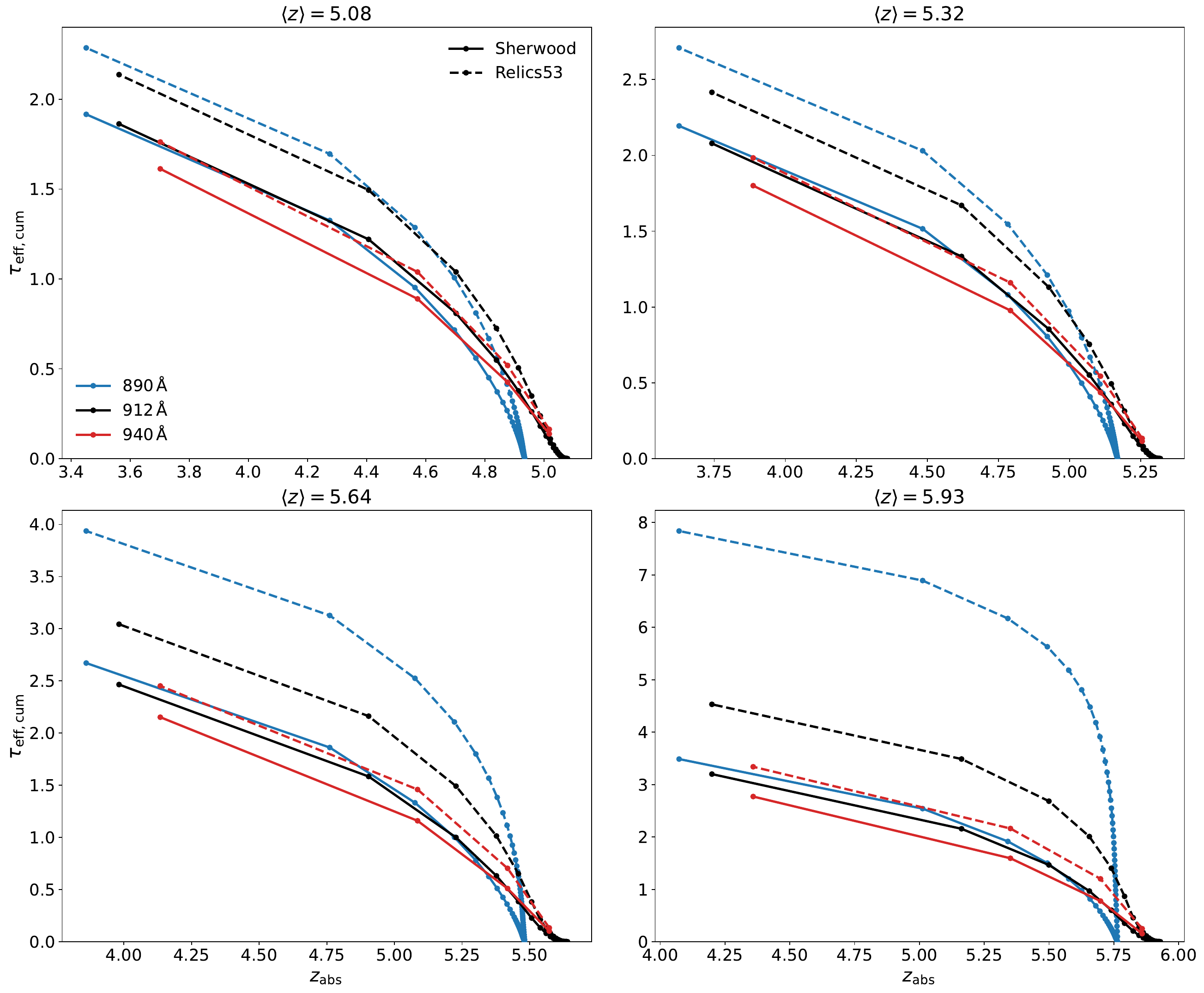}
    \caption{
Same as Fig.~\ref{fig:cumulative_tau_eff_912}, but including $\lambda_{\rm rest}=890$, 912, and 940~\AA. Blue, black, and red curves correspond to these three rest-frame wavelengths, respectively, while solid and dashed lines denote Sherwood and Relics53. The points mark the contribution of individual LyS transitions along each cumulative curve.
}
    \label{fig:cumulative_tau_eff_all}
\end{figure*}

Fig.~\ref{fig:cumulative_tau_eff_all} extends the analysis of
Fig.~\ref{fig:cumulative_tau_eff_912} to $\lambda_{\rm rest}=890$ and
940~\AA, in addition to the reference wavelength of 912~\AA. The
overall behavior is similar at all three wavelengths: the cumulative
LyS effective optical depth builds up through contributions from
multiple transitions, with Relics53 generally producing more absorption
than Sherwood.
The relative contribution of the individual transitions, however,
depends on rest-frame wavelength because the same LyS transition probes
a different absorber redshift at each wavelength. Moving to shorter rest-frame wavelengths, as illustrated by the 890~\AA\ case, the full LyS can contribute, while the absorber redshift associated with a given transition shifts progressively farther from the quasar. In contrast, toward longer wavelengths, as in the 940~\AA\ case, fewer LyS transitions contribute because the highest-order lines increasingly correspond to absorber redshifts beyond the quasar, so their contributions are progressively lost.

\begin{figure*}
    \centering
    \includegraphics[width=\linewidth]{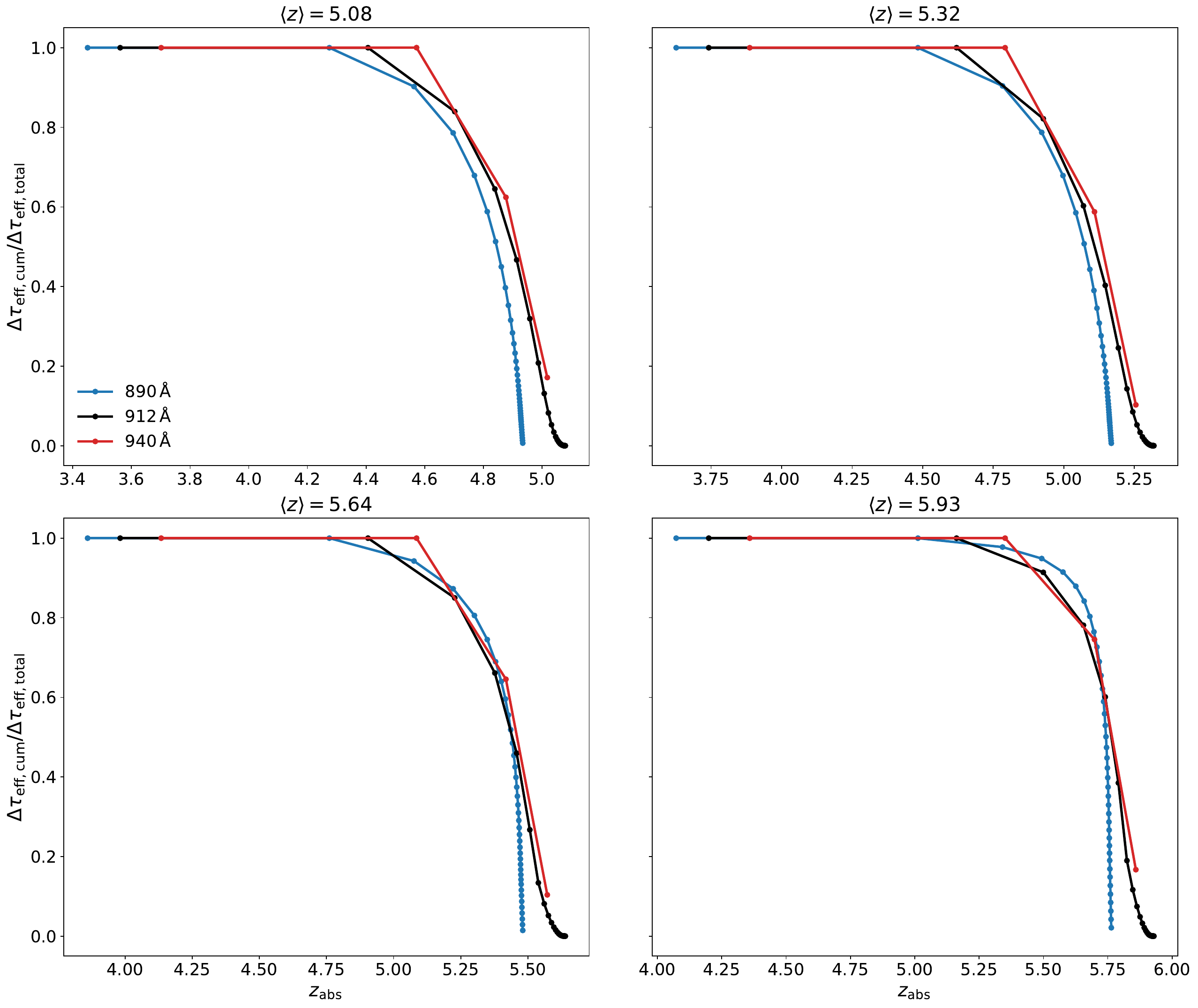}
\caption{
Same as Fig.~\ref{fig:cumulative_delta_tau_eff_912}, but including $\lambda_{\rm rest}=890$, 912, and 940~\AA. Blue, black, and red curves correspond to these three rest-frame wavelengths, respectively. The points mark the contribution of individual LyS transitions along each cumulative curve.
}
    \label{fig:cumulative_delta_tau_eff_all}
\end{figure*}

Fig.~\ref{fig:cumulative_delta_tau_eff_all} shows the fractional cumulative
$\Delta\tau_{\rm eff}$ at $\lambda_{\rm rest}=890$, 912, and 940~\AA.
Toward shorter rest-frame wavelengths, higher-order LyS transitions make
a larger contribution to the cumulative difference, although the
corresponding absorption arises at redshifts progressively farther from
the quasar.

\section{Numerical tests}
\label{sec:Numerical_tests}

\begin{figure*}
    \centering
    \includegraphics[width=\linewidth]{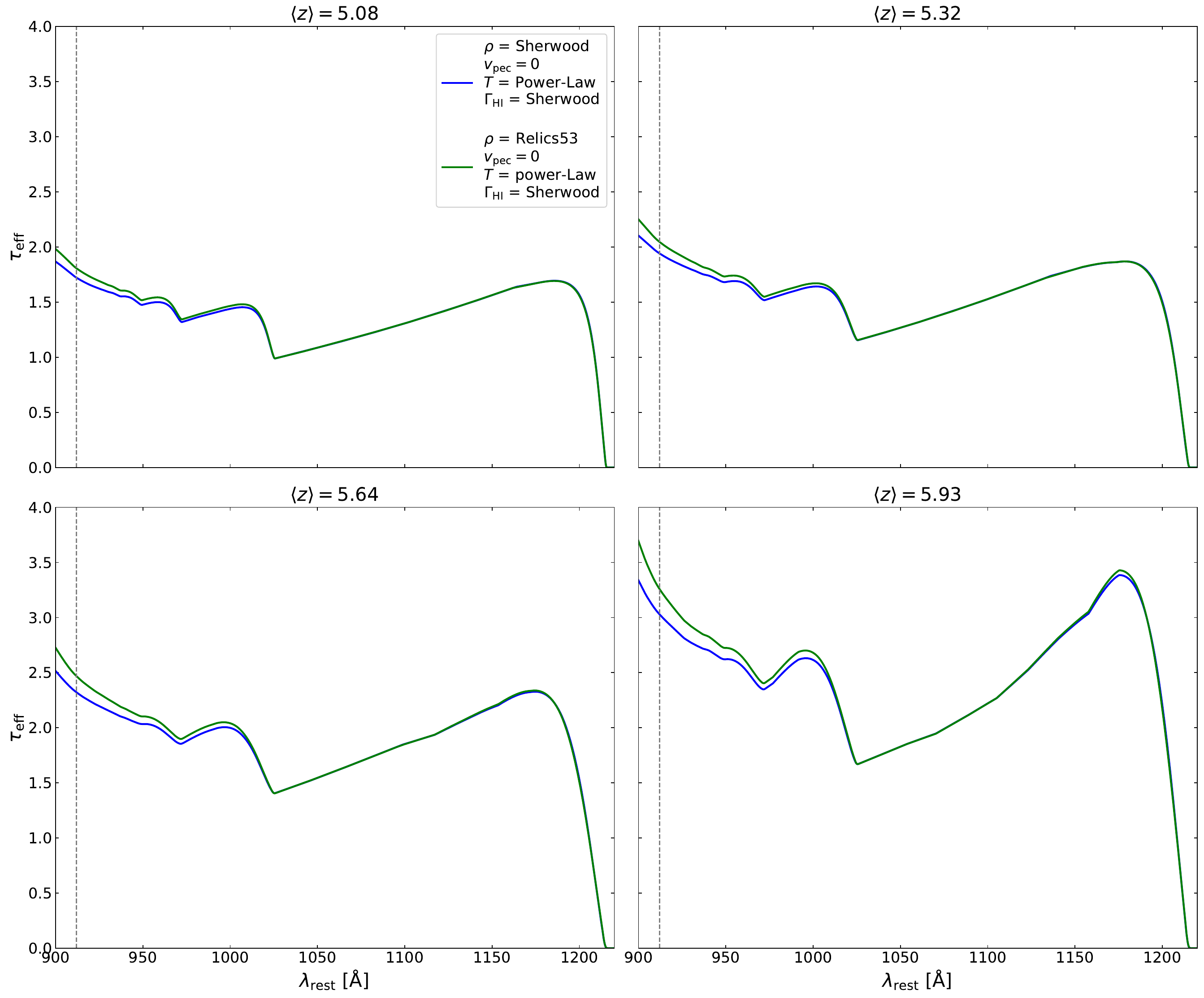}
\caption{Isolating the impact of density on the LyS transmission. Comparison of the Sherwood and Relics53 simulations after setting the peculiar velocities to zero ($v_{\rm pec}=0$), imposing the same power-law $T$--$\rho$ relation, and adopting the Sherwood photoionization rate.
}
    \label{fig:density_test}
\end{figure*}

\begin{figure*}
    \centering
    \includegraphics[width=\linewidth]{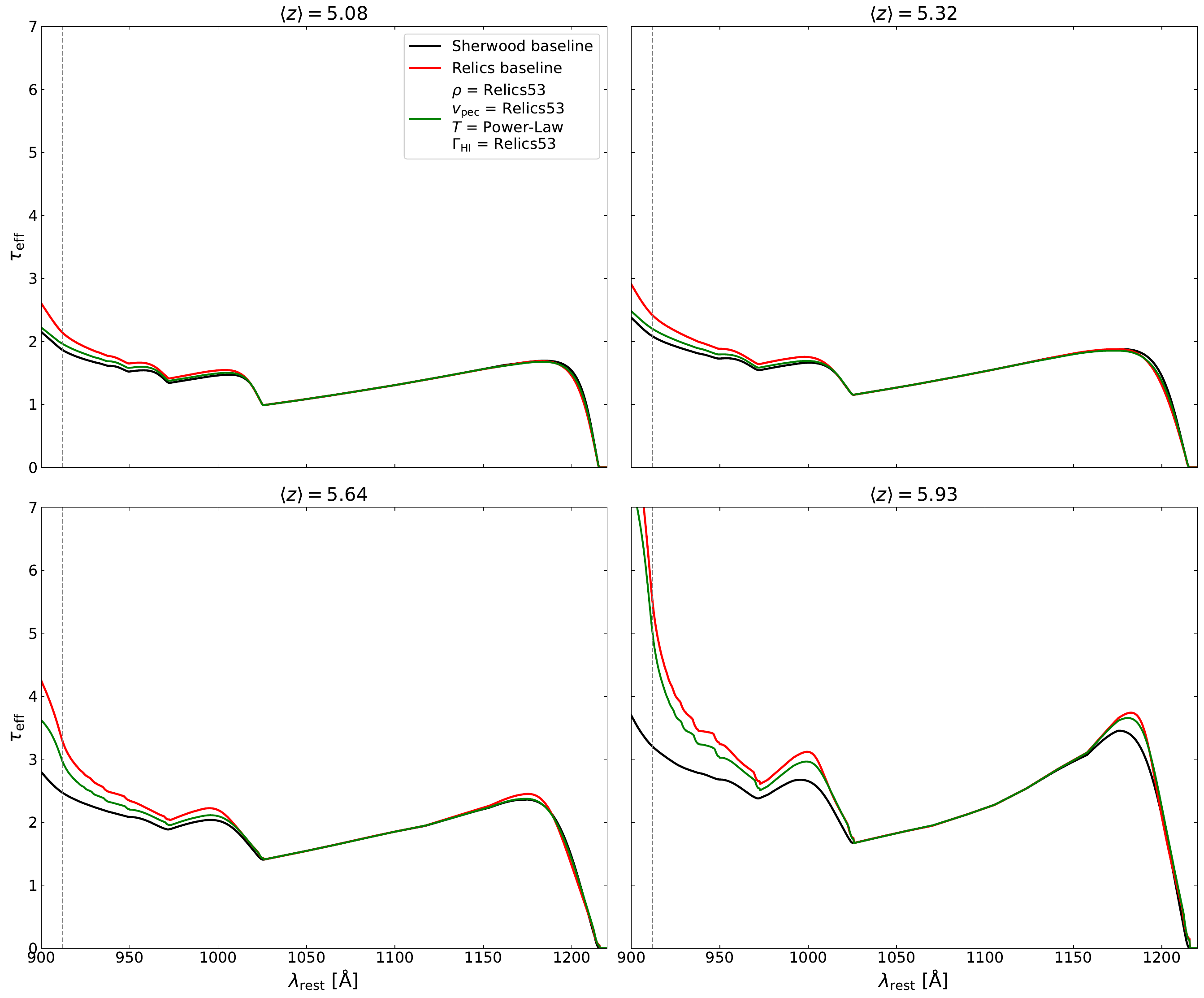}
\caption{
Comparison of the Sherwood and Relics53 simulations after imposing the power-law $T$--$\rho$ relation derived from the Sherwood simulation. The Sherwood simulation is shown only in its original form because its temperature field already closely follows this relation. In contrast, the Relics53 simulation exhibits enhanced transmission, with the magnitude of the change varying with redshift. In these experiments, neutral regions ($x_{\rm HI}>0.1$) are left unchanged and retain their original temperatures.
}
    \label{fig:power_law_test}
\end{figure*}

\begin{figure*}
    \centering
    \includegraphics[width=\linewidth]{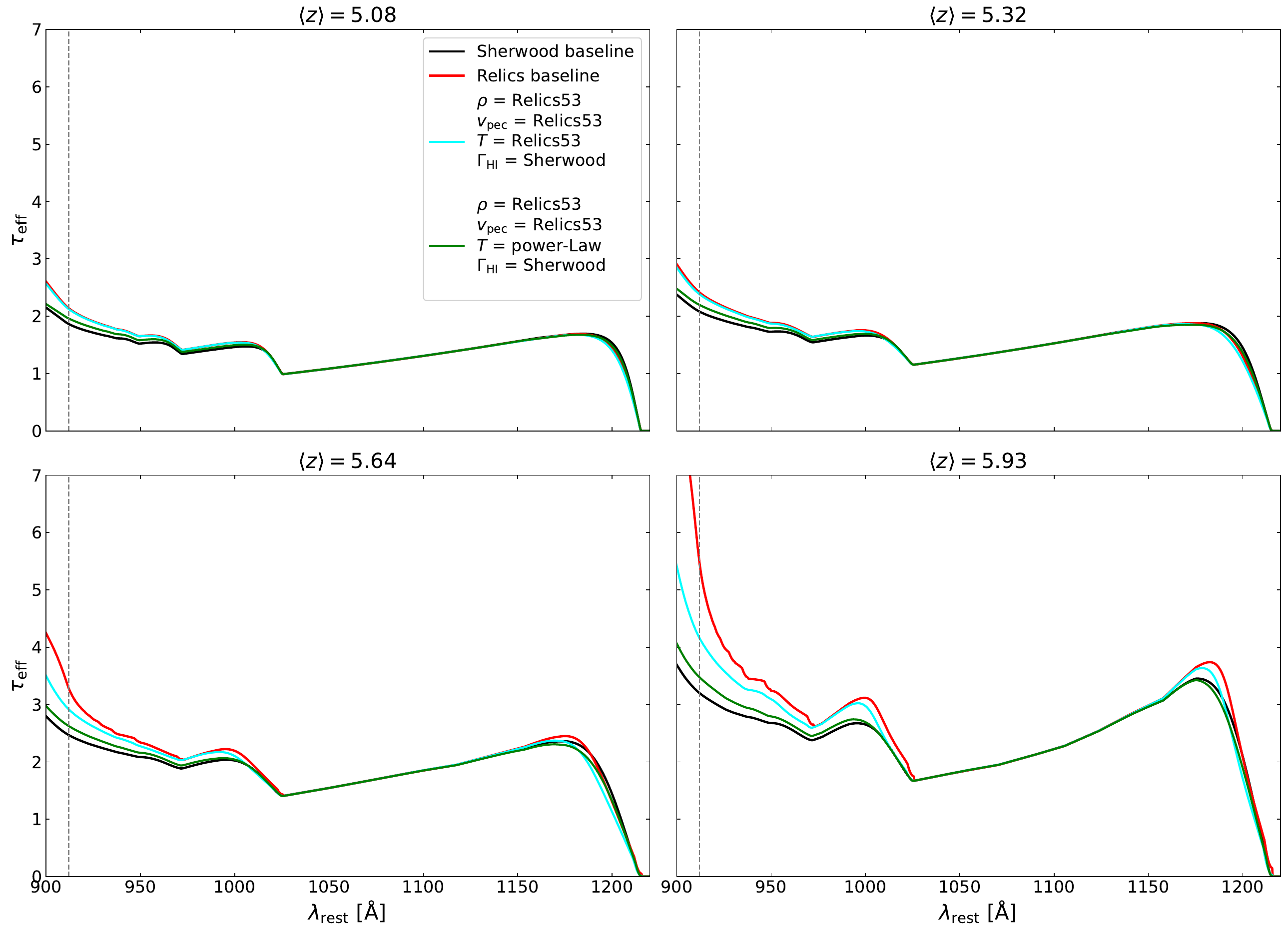}
\caption{
Effect of residual neutral regions on the LyS transmission in Relics53. Converting the neutral regions to ionized while imposing photoionization equilibrium in all regions using a spatially uniform Sherwood photoionization rate (cyan) produces  a significant increase in the predicted transmission for $\langle z \rangle = 5.64$ and 5.93. When the temperatures in all regions are additionally replaced by the Sherwood power-law $T$--$\rho$ relation (green), the transmission shifts further toward the Sherwood prediction.
}
    \label{fig:neutral_region_test}
\end{figure*}

This appendix presents the main controlled numerical experiments used to isolate the physical origin of the differences between the Sherwood and Relics53 simulations. In each experiment, selected physical ingredients are modified while all others are kept fixed, allowing their individual or combined contributions to the predicted LyS transmission to be quantified. The \lya\ optical depths are rescaled to reproduce the observed \lya\ transmission. Figs.~\ref{fig:density_test}, ~\ref{fig:power_law_test}, and ~\ref{fig:neutral_region_test} present the main experiments, whose results are discussed in Section~\ref{subsec:physical_origin}.

To isolate the contribution of the gas density field (Fig.~\ref{fig:density_test}), we set the peculiar velocities to zero in both simulations and impose the same power-law $T$--$\rho$ relation and spatially uniform Sherwood photoionization rate. Under these conditions, differences in the predicted LyS transmission arise from differences in the underlying gas density fields. We then investigate the effect of the thermal state of the ionized IGM (Fig.~\ref{fig:power_law_test}) by replacing the temperatures in ionized regions ($x_{\rm HI}<0.1$) with a power-law $T$--$\rho$ relation based on the Sherwood temperatures, while leaving the temperatures in neutral regions unchanged. Comparing the resulting transmission with the original prediction allows us to assess the impact of departures from the Sherwood power-law $T$--$\rho$ relation in the ionized IGM.

Finally, we investigate the role of residual neutral regions (Fig.~\ref{fig:neutral_region_test}) through two successive modifications to Relics53. First, we impose photoionization equilibrium in all regions using the spatially uniform Sherwood photoionization rate, while retaining the original temperature field. This allows us to assess the effect of differences in the photoionization rate. We then additionally replace the temperatures in all regions, including residual neutral regions, with the Sherwood power-law $T$--$\rho$ relation. Comparing these experiments allows us to assess the additional contribution of the low temperatures in residual neutral regions to the enhanced LyS absorption.

\bsp	
\label{lastpage}

\end{document}